\documentclass[journal]{IEEEtran}

\usepackage{xspace,amsthm,amsmath,amssymb,amsfonts,syntonly}
\usepackage{amssymb} %%%
\usepackage{epsfig} % for postscript graphics files
\usepackage{epstopdf}
\usepackage{epsf,graphics,graphicx,subfig}
\usepackage{cite,bm,color,url,textcomp}
\usepackage{float}
\usepackage{algorithmicx,algpseudocode}
\usepackage{algorithm}
\usepackage{setspace}
\usepackage{cite}
\usepackage{stfloats}
\usepackage{tikz}
\usepackage{lipsum}

\newcommand{\ba}{\begin{array}}
\newcommand{\ea}{\end{array}}
\newcommand{\be}{\begin{displaymath}}
\newcommand{\ee}{\end{displaymath}}
\newcommand{\ben}{\begin{equation}}
\newcommand{\een}{\end{equation}}
\newcommand{\bena}{\begin{eqnarray}}
\newcommand{\eena}{\end{eqnarray}}
\newcommand{\beqa}{\begin{eqnarray*}}
\newcommand{\enqa}{\end{eqnarray*}}

\newcommand{\bc}{\begin{center}}
\newcommand{\ec}{\end{center}}
\newcommand{\bi}{\begin{itemize}}
\newcommand{\ei}{\end{itemize}}
\newcommand{\benu}{\begin{enumerate}}
\newcommand{\eenu}{\end{enumerate}}
\newcommand{\bdes}{\begin{description}}
\newcommand{\edes}{\end{description}}
\newcommand{\bt}{\begin{tabular}}
\newcommand{\et}{\end{tabular}}

\newcommand \alphabf{\mbox{\boldmath$\alpha$\unboldmath}}

\newcommand \omegabf{\mbox{\boldmath$\omega$\unboldmath}}

\newcommand \abf{{\bf a}}

\newcommand \dbf{{\bf d}}
\newcommand \ebf{{\bf e}}

\newcommand \nbf{{\bf n}}

\newcommand \ubf{{\bf u}}

\newcommand \wbf{{\bf w}}
\newcommand \xbf{{\bf x}}
\newcommand \ybf{{\bf y}}
\newcommand \zbf{{\bf z}}

\newcommand \Abf{{\bf A}}

\newcommand \Hbf{{\bf H}}
\newcommand \Ibf{{\bf I}}

\newcommand \Vbf{{\bf V}}

\newcommand{\circlambda}{\mbox{$\Lambda$
             \kern-.85em\raise1.5ex
             \hbox{$\scriptstyle{\circ}$}}\,}

\algdef{SE}[DOWHILE]{Do}{doWhile}{\algorithmicdo}[1]{\algorithmicwhile\ #1}%
\newcommand\blfootnote[1]{%
\begingroup
\renewcommand\thefootnote{}\footnote{#1}%
\addtocounter{footnote}{-1}%
\endgroup
}

\newcommand*{\circled}[1]{\lower.7ex\hbox{\tikz\draw (0pt, 0pt)
circle (.4em) node {\makebox[0.8em][c]{\small #1}};}}
\def\QED{\mbox{\rule[0pt]{1.5ex}{1.5ex}}}

\makeatletter

\newcommand{\Rmnum}[1]{\expandafter\@slowromancap\romannumeral #1@}
\makeatother

\begin{document}
\title{Model-Based Neural Network and Its Application to Line Spectral Estimation}
\author{Yi Jiang*, {\it Member, IEEE}, Tianyi Zhang* and Wei Zhang, {\it Student Member, IEEE}
}
\renewcommand{\thefootnote}{\fnsymbol{footnote}}
\maketitle
\blfootnote{This work was supported by National Natural Science Foundation of China Grant No. 61771005. {(\it Corresponding author: Yi Jiang.)}

The authors are with the Key Laboratory for Information Science of Electromagnetic Waves (MoE), Department of Communication Science and Engineering, School of Information Science and Technology, Fudan University, Shanghai, China (Emails: yijiang@fudan.edu.cn, tianyi\_zhang@fudan.edu.cn, wzhang19@fudan.edu.cn).

* These authors are co-first authors.

The corresponding code can be downloaded from \url{https://github.com/csrlab-fudan/MNN-spectral-estimation}}
\begin{abstract}
This paper presents the concept of ``model-based neural network'' (MNN), which is inspired by the classic artificial neural network (ANN) but for different usages. Instead of being used as a data-driven classifier, a MNN serves as a modeling tool with artfully defined inputs, outputs, and activation functions which have explicit physical meanings. Owing to the same layered form as an ANN, a MNN can also be optimized using the back-propagation (BP) algorithm. As an interesting application, the classic problem of line spectral estimation can be modeled by a MNN. We propose to first initialize the MNN by the fast Fourier transform (FFT) based spectral estimation, and then optimize the MNN by the BP algorithm, which automatically yields the maximum likelihood (ML) parameter estimation of the frequency spectrum. We also design a method of merging and pruning the hidden-layer nodes of the MNN, which can be used for model-order selection, i.e., to estimate the number of sinusoids. Numerical simulations verify the effectiveness of the proposed method.

\end{abstract}
\begin{IEEEkeywords}
Model-based neural network, spectral analysis, back propagation, SPICE, IAA
\end{IEEEkeywords}
\IEEEpeerreviewmaketitle
\section{Introduction}\label{introduction}
As a powerful tool of machine learning, an artificial neural network (ANN) can closely approximate any function with multiple hidden-layer nodes and  nonlinear activation functions \cite{Goodfellow-et-al-2016}. Owing to its capability of universal approximation \cite{HORNIK1991251}, an ANN can be trained as an excellent classifier using the famous back-propagation (BP) algorithm driven by ``big data''; thus, the ANN has in recent years been applied to various fields, like computer vision \cite{LDCT19}, natural language processing \cite{tan2019multilingual} and so on.

Efforts have also been made to use ANN for solving some signal processing problems in communications. For example, in \cite{YLJ18} and \cite{SDW19}, ANN has been used to obtain the channel state information. This kind of methods can have good performance in certain scenarios, but commonly lacks interpretability and generalization capability, probably because the {\it nonlinear} activation functions in the ANN neurons, such as sigmoid, softmax, ReLU, are defined from a mathematical perspective but lack clear physical meanings. For the same reason, an ANN is usually regarded as a data-driven tool rather than a model-based one, despite some recent efforts to combine the ANN with certain model-based domain knowledge as done in \cite{liaozhaogaoli2020,shlezinger2020viterbinet,yangdujiang2021}.

In this paper, we propose the concept of ``model-based neural network'' (MNN), which stems from the first author's previous work in relay network optimization \cite{WangJiang2020}\cite{WangJiang2022} (but termed as {\it quasi-neural network} therein). The MNN has the same layered form as the ANN but with the input, the output, and the activation functions being artfully designed to have explicit physical meanings. Hence, the MNN can be regarded as an evolution of the classic ANN towards a fully interpretable modeling tool. The MNN has the key features as follows.
\begin{itemize}
    \item Different from an ANN as a data-driven classifier, which lacks interpretability and generalization capability \cite{FXLW21, ZLGW03}, the MNN is a modeling tool with clear physical meaning, and hence is fully interpretable;
    \item Similar to an ANN as a universal approximator, the MNN has the layered structure and can be a universal ``modeller'' by artfully choosing the input, the output, and the nonlinear activation functions;
    \item Owing to the layered structure, the MNN can be efficiently optimized using the BP algorithm based on the chain rule of derivative.
%     \item[3)] Owing to the layered structure, the MNN can also be optimized using the back-propagation (BP) algorithm based on the chain rule of derivative; thus, the MNN allows for parallel computation conducted on a graphic processing unit (GPU) ultra-efficiently, which makes it suitable for solving large-scale problems.
\end{itemize}

%The MNN can be applied to solve a multitude of complicated nonlinear problems.
As a showcase application of the MNN,
%3) The model-order selection problem can be solved by  merging and pruning the hidden-layer nodes of the MNN according to some criteria with solid math. %, which essentially selects the order of the model is either unknown or variant (e.g., time-variant channel estimation problem).
% We transform the traditional neural network into a modeling tool by skillfully selecting the activation function and the network weights, and transform the original problem of parametric spectral estimation into the problem of neural network weight optimization.
we apply the MNN to the classic spectral estimation \cite{stoica2005spectral}.

Spectral estimation has been actively researched for several decades, due to its wide applications in radar \cite{YLSXB2010}, medical imaging \cite{LDSP2008}, wireless communications \cite{RB1998}, and autonomous-driving vehicles \cite{SPP20}, etc. % Solving the spectral estimation problem can sufficiently validate the feasibility and superiority of MNN for this signal processing application.
%   like spectral analysis and angle-of-arrival/angle-of-departure (AoA/AoD) estimation.
Three categories of methods have been developed for spectral estimation in the past several decades. The first category is non-parametric, among which the fast Fourier transform (FFT) is the simplest and most widely-used. This category of methods, however, often suffers from low resolution and high false alarm probability.

The second category parameterizes the signal and estimates the parameters based on the maximum likelihood (ML) criterion. The ML estimation is asymptotically statistically efficient under white Gaussian noise if the number of the sinusoids is known and the global optimum is achieved \cite{stoica2005spectral}, but it is computationally very involved due to the non-convexity of the problem. The state-of-the-art methods include the RELAX method \cite{listoica1996efficient,liuli1998implement}, the atomic norm based method \cite{bhaskar2013atomic}\cite{yang2015gridless}, and the Newtonized Orthogonal Matching Pursuit (NOMP) method \cite{MRM16}. The NOMP method also proposes an approach to determine the number of the sinusoids, i.e., the model order. Both RELAX and NOMP methods are essentially coordinate descent methods and hence can be time-consuming when the model order is large. The atomic norm based methods rely on semidefinite programming (SDP) and hence is computationally complicated especially in the high-dimension scenarios. The MUSIC algorithm \cite{R1986} and the ESPRIT algorithm \cite{RoyKailath1989}, as two famous parametric methods, cannot yield real ML estimates even with the known number of sinusoids, and hence are not statistically efficient.

%Multiple iterations are necessary to obtain good performance.
The third category is the semi-parametric methods, an intermediate one between the first and the second categories. This kind of methods utilize the signal sparsity and solve a convex problem. They can often achieve higher resolution and  lower false-alarm probability than the non-parametric methods. But they are not as theoretically robust as the parametric ones, because the asymptotic efficiency is not guaranteed \cite{SBL11,SZL14}. Moreover, the semi-parametric methods are grid-based, leading to performances confined to the granularity of the grid points.

In this paper, we apply the MNN as a new solution to the problem of line spectral estimation. It belongs to the second category, i.e., it is a parametric method. But it can solve the non-convex problem efficiently and outperform the state-of-the-art methods. Specifically, we use the time index as the MNN's input, use the complex amplitudes and digital angular frequencies as the network's weights, and use the complex exponential function as the nonlinear activation function. Based on the cost function of fitting residual, the BP method \cite{Goodfellow-et-al-2016} is then used to train this MNN. We first obtain the coarse initial estimates of the frequencies and the amplitudes using the simple FFT-based spectral estimation method, which usually provides good initialization for the BP algorithm to find a global optimum. The resultant optimized weights of the MNN are nothing but the optimal estimates of the frequencies and the amplitudes of the sinusoids.

% Note that some recently-developed methods, like Newtonized Orthogonal Matching Pursuit (NOMP) \cite{MRM16}, also use optimization methods, i.e., Newton's method, to solve the spectral estimation problem. However, our MNN method is not like NOMP since MNN can calculate all sinusoids in parallel, while NOMP needs to obtain each sinusoids sequentially. Thus, NOMP may suffer from high-computational complexity and cannot be applied by using GPU, which is not suitable for large-scale problems.

The contributions of this paper are summarized as follows:

1) We introduce the concept of MNN as a universal modeler with clear physical meaning, which may motivate future researches beyond the spectral estimation and the relay communications as studied in \cite{WangJiang2022}.

2) We apply the MNN to solve the classic line spectral estimation problem. Detailed network structure and derivation of updating formula are provided in this paper. Numerical examples show the feasibility of MNN and its superior performance over some widely-used spectral estimation methods, such as the FFT, the MUSIC  \cite{R1986}, and the grid-based semi-parametric methods \cite{SBL11,SZL14}.

% 2) Since line spectral estimation is highly non-convex, applying the BP algorithm with random initialization will often lead to local minima. To avoid the local minima, we use the simple FFT-based spectral estimation method to initialize the MNN, which is usually good enough for the BP algorithm to find a global optimum.
%The proposed method has higher probability to find a global optimum than the parametric method without time-consuming exhaustive search.

3) Model-order selection, i.e., to determine the number of sinusoids, is a challenging aspect of spectral estimation. The classic model-order selection methods, such as the AIC \cite{A74} and the BIC \cite{SS2004}, need exhaustive searches across different model orders until finding the best one, which entails formidable computational complexity. For the MNN, each hidden node corresponds to a sinusoid; thus, model-order selection can be easily achieved via merging and pruning of the hidden-layer nodes. We present criteria of node merging and pruning that is theoretically solid.

The rest of this paper is organized as follows. In Section \ref{sec:model}, we introduce the signal model and formulate the problem. In Section \ref{sec:BP}, we apply the MNN to the classic spectral estimation problem and use the BP to train the network. We also show how to use FFT to initialize the MNN and how to determine the number of sinusoidal components in the line spectral signal by merging and pruning the network nodes. In Section \ref{sec:simulation}, we provide numerical simulation results to verify the effectiveness of our proposed method.

\textit{Notation:} We denote vectors and matrices by boldface lower-case and upper-case letter, respectively. $(\cdot)^T$ and $(\cdot)^H$ denote the transpose and conjugate transpose operation, respectively. $\|\xbf\|_2$ denotes the $\ell_2$ norm of the vector $\xbf$. $\odot$ denotes the element-wise product of two matrices or two vectors. $(\cdot)^{T}$, $(\cdot)^{*}$ and $(\cdot)^H$ denote transpose, complex conjugate and conjugate transpose, respectively. $\mathcal{CN}(0, \sigma^2)$ denotes the complex Gaussian noise with zero mean and $\sigma^2$ variance. $\mathcal{E}(1/\theta)$ denotes the exponential distribution with the mean being $\theta$. $\mathcal{X}^2_{\nu}$ is the chi-square distribution whose degree of freedom is $\nu$. $F_{\nu_1, \nu_2}$ denotes the F-distribution with two degrees of freedom parameters being $\nu_1$ and $\nu_2$.

\section{Signal Model and Model-based Neural Network}\label{sec:model}
We first introduce the concept of the MNN using an example of signal processing for relay network communications, before showing that the MNN can also be applied for line spectral estimation.
\subsection{The MNN for Modeling Relay Networks}
%The so-termed MNN has the same layered form as the conventional ANN but with activation functions artfully designed with clear physical meaning.
%\begin{figure}[ht]
%\centering
%\begin{tabular}{c}
%{\psfig{figure=neuralNet.eps,width=3.5in}}
%\end{tabular}
%\caption{A relay network shown in the upper subplot is analogous to a four-layer ANN shown in the lower subplot. \label{fig.neuralNet}}
%\end{figure}

\begin{figure}[htb]
\centering
\includegraphics[width=3.5in]{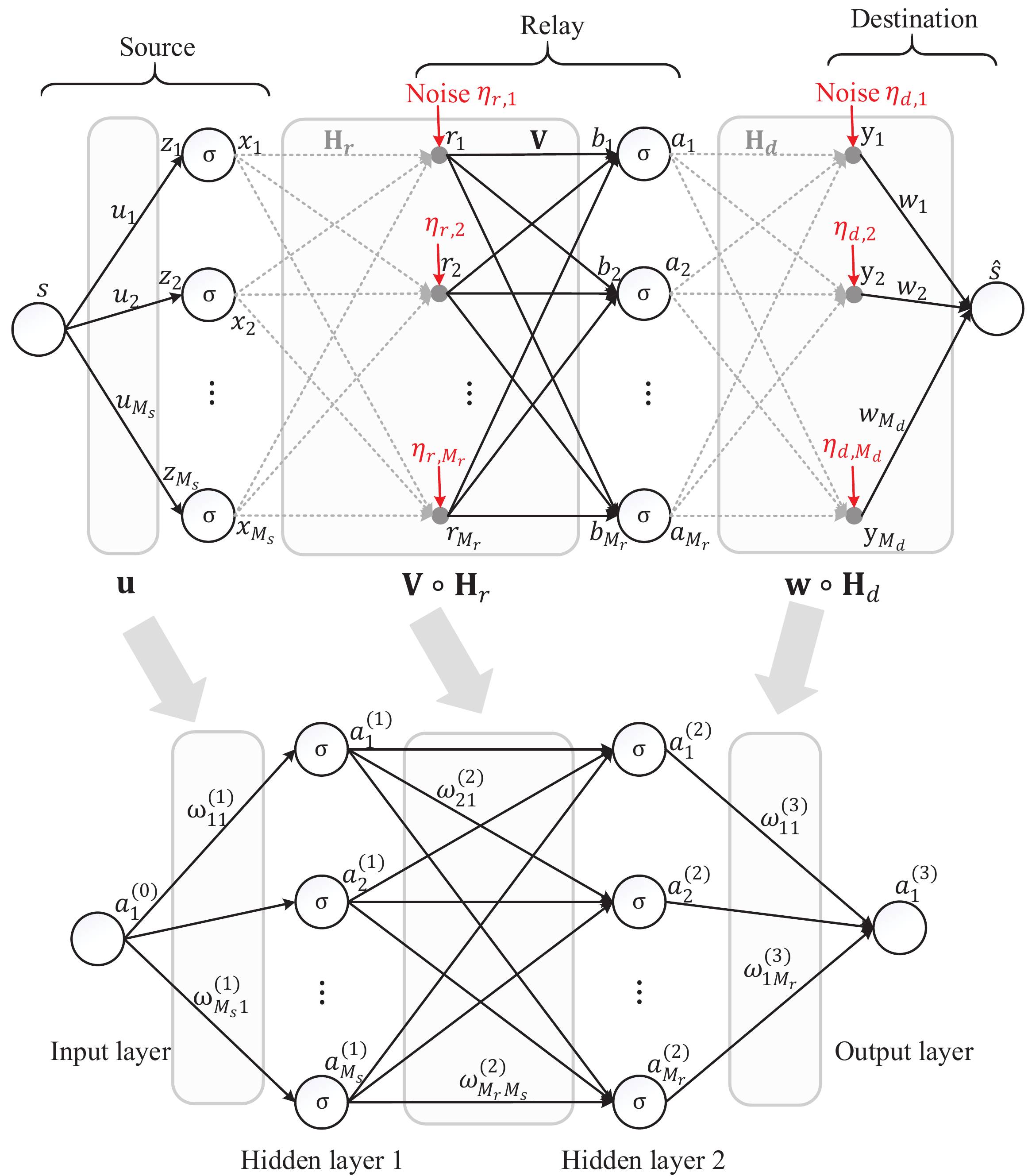}
\caption{A relay network shown in the upper subplot is analogous to a four-layer ANN shown in the lower subplot. }
\label{fig.neuralNet}
\end{figure}

As an illustrative example of the MNN, we recall the nonlinear relay beamforming network studied in \cite{WangJiang2020}.  As shown in the upper subplot of Fig. \ref{fig.neuralNet}, we considered in  \cite{WangJiang2020} the optimization of the precoding of the transmitter, the relay beamforming weights, and the receiver beamforming, denoted by $\ubf$, $\Vbf$, and $\wbf$, respectively, according to the minimum mean squared error (MMSE) criterion. The instantaneous power constraint
per transmit antenna is modeled by the {\it nonlinear} Soft Envelop Limiter (SEL) function % \cite[Equation (38)]{dardari2000a}
\ben\label{eq.sigma}
 \sigma(x)\triangleq
\begin{cases}
x   & {|x|\leq1}\\
e^{j\angle(x)}   & {|x|>1}.
\end{cases}
\een
%to clip the signal for instantaneous power constraint
Then the nonlinear SEL is analogous to a nonlinear activation function of the conventional ANN. Combining $\Vbf$ with the source-to-relay channel (denoted as $\Hbf_r$ in the upper subplot of Fig. \ref{fig.neuralNet})  and combining $\wbf$ with the relay-to-destination channel  (denoted as $\Hbf_d$), we can view the relay network as a four-layer ANN as illustrated in the upper subplot of Fig. \ref{fig.neuralNet}. Such a network has the same layered form as the ANN but with activation functions artfully designed with clear physical meaning. Owing to the layered form, the network can be optimized by the classic BP algorithm based on some pilot sequences.

% Although it was termed as {\it quasi-neural network} in \cite{WangJiang2020}.

% We present the diagram of the relay network in the upper subplot of Fig. \ref{fig.neuralNet}, and observe its striking analogies to the ANN shown in the lower subplot, which are as follows:
% \begin{itemize}
% \item[i)] the antennas of the source, the relay, and the destination are analogous to the neurons in the different layers of the ANN;
% \item[ii)] the data transmission from the source node to the relay node and then to the destination is like the propagation of the training data between the layers in the ANN;
% \item[iii)] the operations $\ubf$, $\Vbf \circ \Hbf_r$, and $\wbf\circ\Hbf_d$ in the relay network are analogous to the connection weights $\omegabf^{(l)}, l=1,2,3$ in the four-layer ANN, respectively, as illustrated by the the gray rectangles in Fig. \ref{fig.neuralNet};
% \item[iv)] the SEL $\sigma(\cdot)$ of the source and the relay is analogous to the activation function of the hidden layer 1 and hidden layer 2 in ANN,  although the latter typically uses a Sigmoid function or a Rectified Linear Unit (ReLU).
% \end{itemize}

\subsection{The MNN for Modeling Line Spectral Signals}
The classic problem of line spectral estimation relies on the signal model\cite{stoica2005spectral}:
\ben
\ybf = \xbf + \ebf \in\mathbb{C}^{N\times 1},
\label{equ.yn}
\een
where $\xbf$ is the sum of $K$ complex-valued sinusoidal signals, i.e., $x(n) = \sum_{k=1}^K\alpha_ke^{j\omega_kn},\ n=0,\dots,N-1$; $\alpha_k$ and $\omega_k\in \left[0,2\pi\right]$ are the complex-valued amplitude and digital angular frequency of the $k$-th complex exponential component, respectively; $\ebf$ is complex i.i.d. additive white Gaussian noise (AWGN) with zero mean and unknown variance $\sigma^2$, i.e., $e(n)\sim\mathcal{CN}(0, \sigma^2)$.

\begin{figure}[htb]
\centering
\includegraphics[width=3.2in]{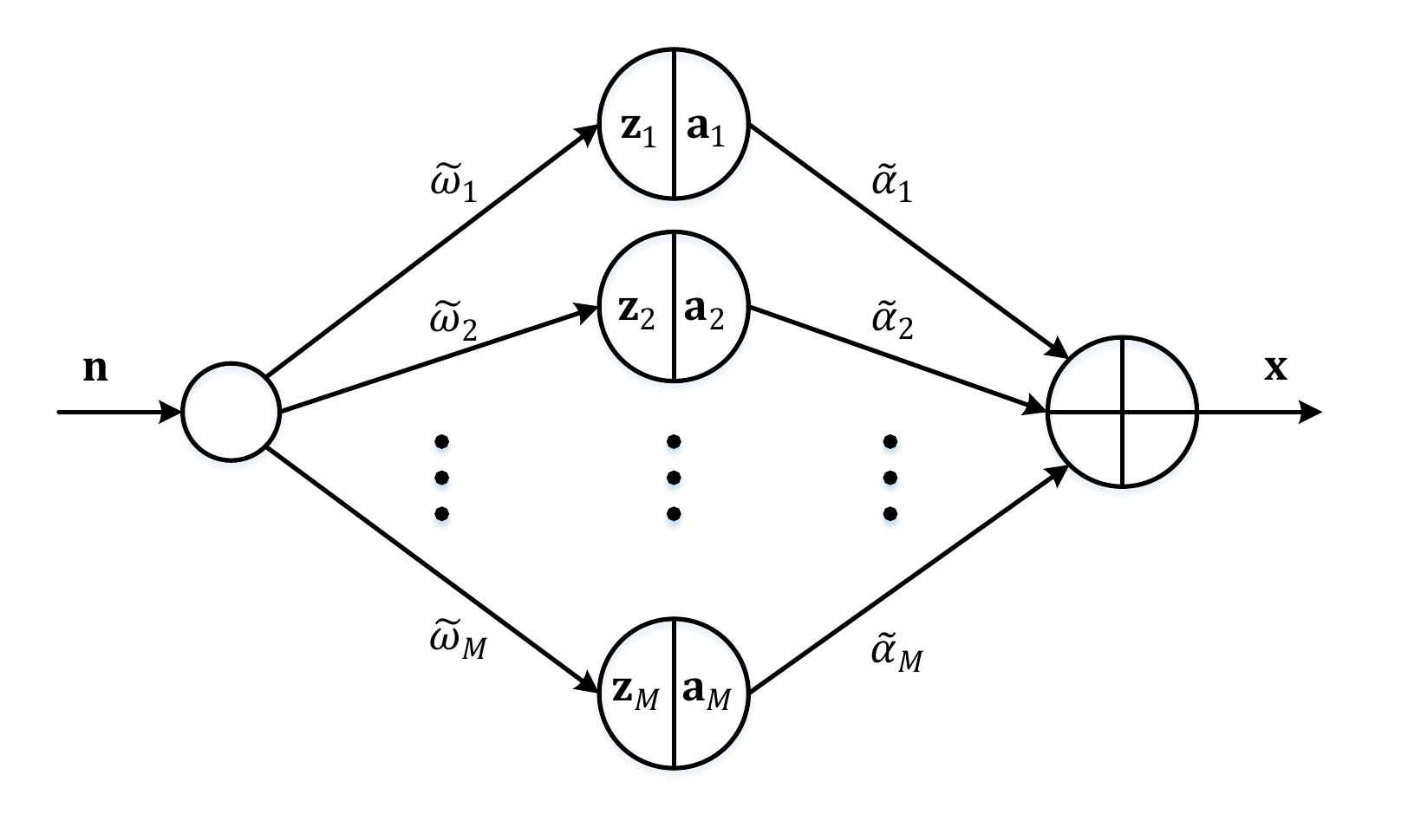}
\caption{The MNN for modeling the superimposed sinousoids.}
\label{fig.NN}
\end{figure}

We propose to model the signal $\xbf$ (\ref{equ.yn}) using a network as shown in Fig. \ref{fig.NN}, where the input is the sequence
\ben
\nbf \triangleq \left[0,1,\dots, N-1 \right]^T\in{\mathbb R}^{N\times1},
\een
the activation function of the  $M$ neurons of the  hidden layer is
\ben \sigma(z) = e^{jz}. \een

Denote $\tilde{\omega}$'s as the weights connecting the input layer and the hidden layer, and $\tilde{\alpha}$'s as the weights connecting the hidden layer and the output layer. Then
we have
\ben
\begin{split}
\zbf_{i} = \tilde{\omega}_i\nbf,\quad &
\abf(\tilde{\omega}_i) = \sigma(\zbf_{i}) = \begin{pmatrix}
e^{jz_{i,1}} \\
e^{jz_{i,2}} \\
\vdots \\
e^{jz_{i,N}}
\end{pmatrix}
=\ \begin{pmatrix}
1 \\
e^{j \tilde{\omega}_i} \\
\vdots \\
e^{j\tilde{\omega}_i(N-1)}
\end{pmatrix}
,%\ i = 1,2,\dots, M,
\label{equ.ziai}
\end{split}
\een
and
\ben
\xbf = \sum_{i=1}^M\tilde{\alpha}_i\abf(\tilde{\omega}_i).
\label{equ.tildeyv1}
\een
Denote $\tilde{\alphabf} = [\tilde{\alpha}_1,\tilde{\alpha}_2,\dots,\tilde{\alpha}_M]^T\in{\mathbb C}^{M\times 1}$, and for notational simplicity, denote $\abf_{i} = \abf(\tilde{\omega}_i)$. Hence,
\ben
\Abf(\tilde{\omegabf}) = [\abf_{1},\abf_{2},\dots,\abf_{M}]\in{\mathbb C}^{N\times M}, \label{equ.Aalpha}
\een
and
\ben
\xbf =  \Abf(\tilde{\omegabf})\tilde{\alphabf}.
\label{equx2}
\een
%$M$ is the number of neurons used in the hidden layer, i.e., the assumed or estimated number of complex sinusoidal components in the signal. $\sigma(\cdot)$ is the activation function of the NN in Figure \ref{fig.NN}. We define the
%and inserting (\ref{equ.ziai}) and (\ref{equ.actfunc}) into (\ref{equ.tildeyv1}),
%we can obtain
%\ben
%\xbf = \sum_{i=1}^M\tilde{\alpha}_ie^{j\tilde{\omega}_in}.
%\label{equ.tildeyv2}
%\een

Note that $\xbf$ in (\ref{equx2}) is the same with that in (\ref{equ.yn}) except that $K$ is replaced by $M$, since the number of signals $K$ is usually unknown in practice. Indeed, the estimation of the model order is challenging, which will be addressed in Section \ref{sec:PM}. % Let us assume that $M\ge K$ for now.

To estimate $\alpha_k$ and $\omega_k$, we choose to adopt the cost function % $\sum_{n=0}^{N-1}\left|y(n)-\sum_{k=1}^K\alpha_ke^{j\omega_kn}\right|^{2}$, i.e.,
% \begin{align}
% &\nonumber \mathop{\min}_{\{\alpha_k,\omega_k\}_k^K}  \quad {\sum_{n=0}^{N-1}\left|y(n)-\sum_{k=1}^K\alpha_ke^{j\omega_kn}\right|^{2}} \\
% &\text{subject to} \quad \omega_k \in \left[0, 2\pi\right]. \label{costFunc}
% \end{align}

% We can choose the cost function
\ben
C(\tilde{\omegabf},\tilde{\alphabf}) \triangleq ||\ybf-\Abf(\tilde{\omegabf})\tilde{\alphabf}||_2^{2},
\label{equ.NNCostFunc}
\een
to train the network. Once the training process converges to the global optimum, the weights, i.e., $\tilde{\omega}_i,\ \tilde{\alpha}_i,\ i= 1,2,\dots,M,$ are naturally the ML estimate of the signal model parameters and contain the complete information of the spectrum of $\ybf$.

The network is similar to a three-layer ANN (with one-hidden layer) in its layered form; thus, the MNNs in both Fig. \ref{fig.neuralNet} and \ref{fig.NN} can be optimized by the BP algorithm. But the MNN differs from a conventional ANN in that the weights and the activation functions of the MNN have perfect physical meaning; thus, the MNN can serve as a modeling tool rather than a data-driven classifier.

% We show in the next section how to achieve line spectral estimation.

\section{Line Spectral Estimation Using MNN} \label{sec:BP}
\subsection{Network Optimization Using BP Algorithm}
To train the MNN, we calculate the gradients of (\ref{equ.NNCostFunc}) with respect to $\tilde{\omega}_i,\tilde{\alpha}_i$ using the BP algorithm, which is essentially a gradient descent method explained as follows.
%\ben
%\frac{\partial C}{\partial \tilde{\alphabf}^*}, \quad \frac{\partial C}{\partial \tilde{\omegabf}},
%\een
%where
%\ben
%\tilde{\omegabf} = [\tilde{\omega}_1,\tilde{\omega}_2,\dots,\tilde{\omega}_M]^T\in{\mathbb R}^{M\times 1}.
%\een

First, by using the chain rule we have from (\ref{equ.NNCostFunc}) that
\ben
\frac{\partial C}{\partial \tilde{\alphabf}^*} = \frac{\partial \xbf}{\partial \tilde{\alphabf}^*}\frac{\partial C}{\partial \xbf}+\frac{\partial \xbf^*}{\partial \tilde{\alphabf}^*}\frac{\partial C}{\partial \xbf^*}.
\een
It follows from (\ref{equ.tildeyv1}) that
\ben
\frac{\partial \xbf}{\partial \tilde{\alphabf}^*} = {\bf 0} \;\; {\rm and} \;\; \frac{\partial \xbf^*}{\partial \tilde{\alphabf}^*} = \Abf^H;
\een
it follows from (\ref{equ.NNCostFunc}) that
\ben
\frac{\partial C}{\partial \xbf} = (\xbf - \ybf)^* \;\; {\rm and} \;\;  \frac{\partial C}{\partial \xbf^*} = (\xbf - \ybf).
\een
Thus, we obtain
\ben
\frac{\partial C}{\partial \tilde{\alphabf}^*} = \Abf^H(\xbf - \ybf).
\label{equ.deralpha}
\een

Second, we have
\ben
\begin{split}
\frac{\partial C}{\partial \tilde{\omega}_i}
=& \left[\frac{\partial \abf_i^*}{\partial \tilde{\omega}_i}\frac{\partial \xbf}{\partial \abf_i^*} + \frac{\partial \abf_i}{\partial \tilde{\omega}_i}\frac{\partial \xbf}{\partial \abf_i}\right]\frac{\partial C}{\partial \xbf} \\
&+ \left[\frac{\partial \abf_i^*}{\partial \tilde{\omega}_i}\frac{\partial \xbf^*}{\partial \abf_i^*} + \frac{\partial \abf_i}{\partial \tilde{\omega}_i}\frac{\partial \xbf^*}{\partial \abf_i}\right]\frac{\partial C}{\partial \xbf^*},\ i = 0,1,\dots,M.
\end{split}
\label{equ.dertildeomegav1}
\een
Knowing from (\ref{equ.tildeyv1}) that
\ben
\frac{\partial \xbf^*}{\partial \abf_i} = {\bf 0}_M\;\; {\rm and} \; \quad \frac{\partial \xbf}{\partial \abf_i^*} = {\bf 0}_M,
\een
we can rewrite (\ref{equ.dertildeomegav1}) as
\ben
\frac{\partial C}{\partial \tilde{\omega}_i}
= \frac{\partial \abf_i}{\partial \tilde{\omega}_i}\frac{\partial \xbf}{\partial \abf_i}\frac{\partial C}{\partial \xbf}
+ \frac{\partial \abf_i^*}{\partial \tilde{\omega}_i}\frac{\partial \xbf^*}{\partial \abf_i^*}\frac{\partial C}{\partial \xbf^*}.
\label{equ.dertildeomegav2}
\een
From (\ref{equ.ziai}), (\ref{equ.tildeyv1}), we can obtain
\ben \label{eqaw}
\begin{split}
\frac{\partial \abf_i}{\partial \tilde{\omega}_i} &= \nbf^T\odot
\left[\frac{\partial a_{i,1}}{\partial z_{i,1}},\frac{\partial a_{i,2}}{\partial z_{i,2}},\dots,\frac{\partial a_{i,N}}{\partial z_{i,N}}\right], \\
\frac{\partial \abf_i^*}{\partial \tilde{\omega}_i} &= \nbf^T\odot
\left[\frac{\partial a^*_{i,1}}{\partial z_{i,1}},\frac{\partial a^*_{i,2}}{\partial z_{i,2}},\dots,\frac{\partial a^*_{i,N}}{\partial z_{i,N}}\right],
\end{split}
\een
and
\ben \label{eqxa}
\frac{\partial \xbf}{\partial \abf_i} = \tilde{\alpha}_i\Ibf_N, \frac{\partial \xbf^*}{\partial \abf_i^*} = \tilde{\alpha}^*_i\Ibf_N,
\een
%\ben
%\frac{\partial C}{\partial \xbf} = (\xbf - \ybf)^*,
%\een
where
\ben
\begin{split}
\frac{\partial a_{i,n}}{\partial z_{i,n}} = je^{jz_{i,n}},\quad &\frac{\partial a_{i,n}^*}{\partial z_{i,n}} = -je^{-jz_{i,n}}, \\
&n = 0,1,\dots,N-1.
\end{split}
\een
Thus, substituting (\ref{equ.deralpha}) (\ref{eqaw}) (\ref{eqxa}) into (\ref{equ.dertildeomegav2}) yields
\ben
\frac{\partial C}{\partial \tilde{\omegabf}} = 2{\rm Im}\left\{\tilde{\alphabf}\odot\left[\Abf^T\left[\nbf\odot(\ybf - \xbf)^*\right]\right]\right\}.
\label{equ.deromega}
\een

Given the gradients (\ref{equ.deralpha}) and (\ref{equ.deromega}), we then use the momentum method \cite{Goodfellow-et-al-2016} to choose the search direction and the learning rate, since it usually outperforms the method of steepest descent, especially for a non-convex problem. In the $t$-th iteration, the network weights are updated as
\ben
\begin{split}
\tilde{\alphabf}(t) &= \tilde{\alphabf}(t-1) - \gamma\dbf_{\tilde{\alphabf}}(t), \\
\tilde{\omegabf}(t) &= \tilde{\omegabf}(t-1) - \gamma\dbf_{\tilde{\omegabf}}(t),
\end{split}
\label{equ.alphaOmegaUpdate}
\een
where $\gamma$ is the learning rate, $\dbf_{\tilde{\alphabf}}(t)$ and $\dbf_{\tilde{\omegabf}}(t)$ are the momentums defined as
\ben
\begin{split}
\dbf_{\tilde{\alphabf}}(t) &= \lambda\dbf_{\tilde{\alphabf}}(t-1)+(1-\lambda)\frac{\partial C}{\partial \tilde{\alphabf}^*}(t), \\
\dbf_{\tilde{\omegabf}}(t) &= \lambda\dbf_{\tilde{\omegabf}}(t-1)+(1-\lambda)\frac{\partial C}{\partial \tilde{\omegabf}}(t),
\end{split}
\label{equ.momentum}
\een
with $\dbf_{\tilde{\alphabf}}(0)={\bf 0}_M$, $\dbf_{\tilde{\omegabf}}(0) = {\bf 0}_M$. Here $\lambda$ is the momentum parameter.

Note that $\tilde{\omega}_i,\ i = 1,2,\dots,M$ obtained by BP algorithm are not necessarily confined to $\left[0, 2\pi\right]$, which is fine because at the end we can simply take the $2\pi$ modulo of $\tilde{\omega}_i$, i.e., $\tilde{\omega}_i \leftarrow \text{mod}(\tilde{\omega}_i, 2\pi),\ i = 1,2,\dots, M$.

The initialization of $\tilde{\alphabf}(0)$ and $\tilde{\omegabf}(0)$ is explained in the next.

\subsection{Initialization Using FFT}\label{sec.initial}
Due to the non-convexity of (\ref{equ.NNCostFunc}), a random initialization of the weight $\tilde{\alphabf}(0)$ and $\tilde{\omegabf}(0)$ often leads to a local optimum; thus, it may require too many random initializations for the BP algorithm before finding a global optimum. To solve this issue, we consider using the FFT to obtain the initial parameter estimation.

First, apply a zero-padded FFT to the sequence $\ybf$ to obtain an $L$-point frequency-domain sequence $\ybf^f$; second, locate of the peaks of $|\ybf^f|$ and add the corresponding frequency points into the initial frequency set $\tilde{\omegabf}(0) \in {\mathbb R}^P$, where $P$ is the number of peaks. Due to the low resolution of the FFT spectrum, one peak may be due to two or more sinusoids with frequencies approximate to each other. To obtain a higher frequency resolution, we also check the frequency points adjacent to the peaks $\tilde{\omega}_i, i=1,...,P$, i.e., to compare the FFT power spectrum at frequencies $\tilde{\omega}_i \pm \frac{2\pi}{L}$ and augment to the vector $\tilde{\omegabf}(0)$ by $\tilde{\omega}_i + \frac{2\pi}{L}$ or $\tilde{\omega}_i - \frac{2\pi}{L}$ depending which one corresponds to the higher power. After removing the repeated elements, the cardinality of $\tilde{\omegabf}(0)$ is denoted by $M$. Finally, $\tilde{\alphabf}$ can be initialized by using the least squared method corresponding to the frequency points in $\tilde{\omegabf}(0)$, i.e.,
\ben\label{equ:a0}
\tilde{\alphabf}(0) = [\Abf^H(\tilde{\omegabf}(0))\Abf(\tilde{\omegabf}(0))]^{-1}\Abf^H(\tilde{\omegabf}(0)) \ybf.
\een

\subsection{Model Order Selection}\label{sec:PM}
As mentioned earlier, it is a nontrivial task to determine the number of sinusoids. As each node in the hidden layer of the MNN corresponds to a sinusoid, we can merge or prune the nodes to adjust the order of the model conveniently when conducting the BP algorithm. The guidance of model order selection is explained as follows.

\subsubsection{Criterion of nodes merging}
To determine whether two nodes should be merged is essentially a hypothesis testing problem:
\ben
\begin{aligned}
    &H_0: \omega_j - \omega_j \leq \Delta\omega_{\min}\\
    &H_1: \omega_j - \omega_i > \Delta\omega_{\min},
\end{aligned}
\een
with $\omega_j>\omega_i$. Here $\Delta\omega_{\min}\geq 0$ is some prescribed number.

To solve this problem, we derive the posterior probability of $\Delta\omega_{ij}\triangleq\omega_j-\omega_i$ conditioned on the ML estimate $\tilde{\omega}_i,\tilde{\omega}_j$, i.e.,
\begin{equation}
    {\rm Pr}(\Delta\omega_{ij}|\tilde{\omega}_i,\tilde{\omega}_j).
\end{equation}

We first use the Cram\'er–Rao bound (CRB) to obtain the probability of the ML estimate. Consider the simplifying scenario where only two sinusoids exist, i.e.,
\begin{equation}\label{equ.2sin}
\begin{split}
    & y(n) = {\alpha}_ie^{j{\omega}_in} + {\alpha}_je^{j{\omega}_jn} + e(n), \\
    & \hspace{2em}{\omega}_i<{\omega}_j, n = 0, \dots, N-1, e(n)\sim\mathcal{CN}(0, \hat{\sigma}^2).
\end{split}
\end{equation}
Then the ML estimates of both frequencies should be unbiased with variance approaching the CRB \cite{ME83}. That is,
\ben
\left.\begin{pmatrix}
\tilde{\omega}_i\\
\tilde{\omega}_j
\end{pmatrix}\right|\begin{pmatrix}
{\omega}_i\\
{\omega}_j
\end{pmatrix} \sim \mathcal{N}\left(\begin{pmatrix}
{\omega}_i\\
{\omega}_j
\end{pmatrix},{\rm CRB}^{ij}\right),
\een
where
\ben\label{equ.CRBomega}
\begin{split}
   &{\rm CRB}^{ij} = \frac{\sigma^2}{2}\frac{1}{|\alpha_i|^2|\alpha_j|^2\rho_1^2 - {\rm Re}\left[\alpha_i^*\alpha_j\rho_2\right]^2}\times \\
   &\hspace{1em}\begin{bmatrix}
   |\alpha_j|^2\rho_1 & -{\rm Re}\left[\alpha_i^*\alpha_j\rho_2\right]\\
   -{\rm Re}\left[\alpha_i^*\alpha_j\rho_2\right]& |\alpha_i|^2\rho_1
   \end{bmatrix},
\end{split}
\een
with $\rho_1$ and $\rho_2$ being defined as
\begin{equation}
%    \begin{split}
        \rho_1 = \sum_{n=0}^{N-1}n^2, \quad
        \rho_2 = \sum_{n=0}^{N-1}n^2e^{j(\omega_j - \omega_i)n}.
    %\end{split}
\end{equation}
The derivation for (\ref{equ.CRBomega}) is relegated to Appendix.

Next, we assume without loss of generality that $\omega_i, \omega_j$ are independent and have a prior probability of uniform distribution in $[0,2\pi]$, % since the sinusoidal function has the period of $2\pi$ and thus, $\omega$ and $\omega+2\pi$ will lead to the same sinusoid. , we assume and obey the uniform distribution on the real number domain, i.e.,
\begin{equation}
    p\left(\begin{pmatrix}
{\omega}_i\\
{\omega}_j
\end{pmatrix}\right) = \frac{1}{4\pi ^2}, \omega_i,\omega_j\in[0,2\pi ].
\end{equation}
Then the posterior distribution of $\omega_i, \omega_j$ can be obtained by
\begin{equation}
\begin{split}
    &p\left(\left.\begin{pmatrix}
{\omega}_i\\
{\omega}_j
\end{pmatrix}\right|\begin{pmatrix}
\tilde{\omega}_i\\
\tilde{\omega}_j
\end{pmatrix}\right) \\
&= \frac{p\left(\left.\begin{pmatrix}
\tilde{\omega}_i\\
\tilde{\omega}_j
\end{pmatrix}\right|\begin{pmatrix}
{\omega}_i\\
{\omega}_j
\end{pmatrix}\right)p\left(\begin{pmatrix}
{\omega}_i\\
{\omega}_j
\end{pmatrix}\right)}{\displaystyle\int_0^{2\pi}\int_0^{2\pi} p\left(\left.\begin{pmatrix}
\tilde{\omega}_i\\
\tilde{\omega}_j
\end{pmatrix}\right|\begin{pmatrix}
{\omega}_i\\
{\omega}_j
\end{pmatrix}\right)p\left(\begin{pmatrix}
{\omega}_i\\
{\omega}_j
\end{pmatrix}\right)d{\omega}_id{\omega}_j}\\
&=p\left(\left.\begin{pmatrix}
\tilde{\omega}_i\\
\tilde{\omega}_j
\end{pmatrix}\right|\begin{pmatrix}
{\omega}_i\\
{\omega}_j
\end{pmatrix}\right).
\end{split}
\end{equation}
Hence, the posteriori distribution of the frequencies conditioned on the ML estimates is
\begin{equation}
   \left.\begin{pmatrix}
    {\omega}_i\\
    {\omega}_j
    \end{pmatrix}
    \right|\begin{pmatrix}
    \tilde{\omega}_i\\
    \tilde{\omega}_j
    \end{pmatrix}\sim \mathcal{N}\left(\begin{pmatrix}
    \tilde{\omega}_i\\
    \tilde{\omega}_j
    \end{pmatrix},{\rm CRB}^{ij}\right).
\end{equation}

It follows from (\ref{equ.CRBomega}) that the statistic $\Delta{\omega}_{ij} \triangleq {\omega}_{j}-{\omega}_{i}$
\ben
\Delta{\omega}_{ij} \sim \mathcal{N}(\tilde{\omega}_{j}-\tilde{\omega}_{i}, {\rm CRB}^{ij}_{\Delta}),
\een
where
\begin{equation}\label{equ.CRBdomega}
    \begin{split}
        {\rm CRB}^{ij}_{\Delta} &= \begin{bmatrix}-1, 1\end{bmatrix}{\rm CRB}^{ij}\begin{bmatrix}-1 \\ 1\end{bmatrix} \\
        & = \frac{\sigma^2}{2}\frac{(|\alpha_i|^2+|\alpha_j|^2)\rho_1 + 2{\rm Re}[\alpha_i^*\alpha_j\rho_2]}{|\alpha_i|^2|\alpha_j|^2\rho_1^2 - {\rm Re}[\alpha_i^*\alpha_j\rho_2]^2}.
    \end{split}
\end{equation}

\begin{figure}[htbp]
  \centering
  \includegraphics[width=0.45\textwidth]{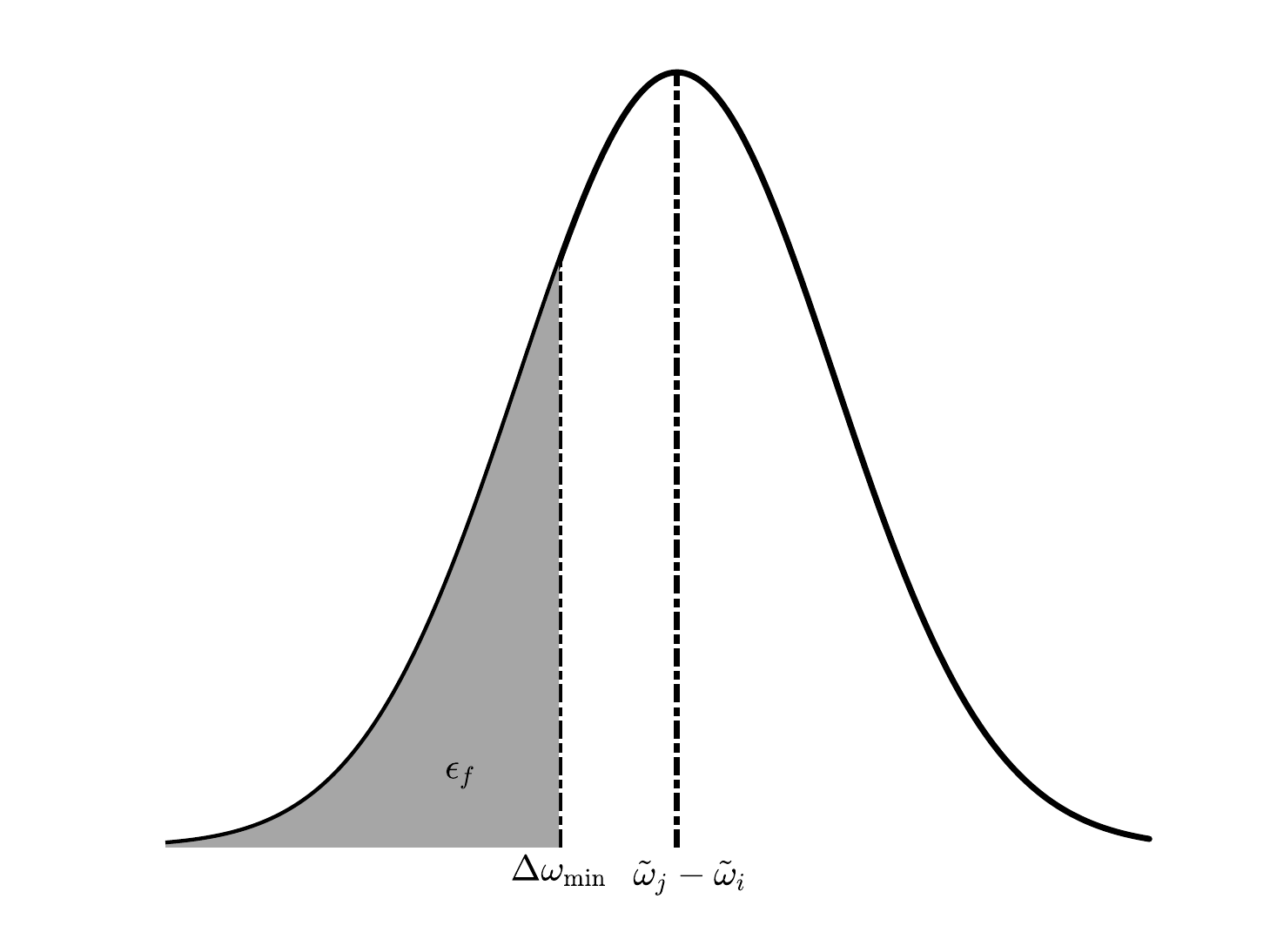}
  \caption{Illustration of determining merging criterion.}
  \label{fig.merge}
\end{figure}

% We set $\Delta\omega_{\min}=0$.
If the probability of $\Delta{\omega}_{ij}\leq \Delta\omega_{\min}$ is larger than a small value $\epsilon_f$ (e.g., $1\times10^{-6}$), i.e.,
\begin{equation}\label{equ:Prmerge}
    {\rm Pr}(\Delta{\omega}_{ij} \leq \Delta\omega_{\min} )> \epsilon_f,
\end{equation}
we accept the hypothesis $H_0$ and propose to merge the two ``hidden'' nodes since they can not be separated correctly with high probability (see Fig. \ref{fig.merge}).
By some simple calculations, we find that if
\begin{equation}\label{equ:merge}
    \tilde{\omega}_j - \tilde{\omega}_i < \Delta\omega_{\min} -\sqrt{{\rm CRB}^{ij}_{\Delta}}\mathcal{N}^{-1}(\epsilon_f),
\end{equation}
where $\mathcal{N}^{-1}(\cdot)$ denotes the inverse function of the cumulative distribution function (CDF) of the standarded Gaussian distribution, we can merge the two corresponding nodes into one, average their frequencies to be $(\tilde{\omega}_i+\tilde{\omega}_j)/2$, and combine the amplitudes into $\tilde{\alpha}_i + \tilde{\alpha}_j$ before the next iteration.

Note that although in real applications we do not know the true value of the complex amplitudes, frequencies, and the Gaussian noise variance, we can use the estimated value $\tilde{\alpha}_i$, $\tilde{\alpha}_j$, $\tilde{\omega}_i$, $\tilde{\omega}_j$ and $\hat{\sigma}^2 = \|\xbf - \ybf\|_2^2/N$ instead of the true value to calculate ${\rm CRB}^{ij}_{\Delta}$, since the ML estimate of the MNN is unbiased. Also, we commonly set $\Delta\omega_{\min}=0$ in real applications for simplicity without loss of performance.

\subsubsection{Criterion of node pruning}\label{sec.prunecriterion}
To determine whether a node should be pruned is also a hypothesis testing problem:
\ben
\begin{aligned}
    &H_0: \mbox{a sinusoid does not exist at } \tilde{\omega} \\
    &H_1: \mbox{a sinusoid exists at } \tilde{\omega} .
\end{aligned}
\een
The idea of deriving the node pruning criterion is that if the power of a sinusoidal component is larger than a certain threshold, we keep this component or it should be pruned.

Consider the statistic
\begin{equation}\label{equ:prune_stat}
    \xi \triangleq \frac{|\abf^H(\tilde{\omega}) \ybf|^2}{\|\ybf - \Abf(\tilde{\omegabf}) {\tilde\alphabf}\|_2^2},
\end{equation}
where $\tilde{\omegabf}$ and ${\tilde\alphabf}$ are the current weights of the MNN. This statistic describes the distribution of the power at frequency $\tilde{\omega}_i$ normalized by the noise power. We propose to prune the node from the network if $\xi$ is less than some threshold $\Xi$ after some iterations. We show in the next that the distribution function of $\xi$ under $H_0$ is independent of the noise power; thus, the threshold $\Xi$ can be derived according to a prescribed constant false alarm rate (CFAR).

With this statistic, we can use the following false alarm rate criterion:
\begin{equation}
    {\rm Pr}\left(\frac{|\abf^H_i\ybf|^2}{\|\ybf - \Abf{\tilde\alphabf}\|_2^2}>\Xi\right)<\epsilon_a, i = 1, \dots, M.
\end{equation}
It means that when $\ybf$ does not contain a sinusoid at frequency $\tilde{\omega}_i$, the probability of false alarm, i.e., the corresponding statistic (\ref{equ:prune_stat}) being larger than a certain threshold $\Xi$, should be less than a small value $\epsilon_a$, e.g., $1\times 10^{-6}$.

To find $\Xi$, we need to derive the distribution of (\ref{equ:prune_stat}). For simplicity, we consider the case that $\ybf$ does not contain any sinusoid, i.e., $\ybf = \ebf \sim \mathcal{CN}(0, \sigma^2)$. We first derive the distributions of the numerator and denominator of (\ref{equ:prune_stat}) separately, and find that they have forms of chi-square distribution. Thus, a scaled version of (\ref{equ:prune_stat}) obeys an F-distribution.

Because $\ybf = \ebf \sim \mathcal{CN}(0, \sigma^2)$, $|y(n)|^2$ obeys the exponential distribution whose mean is $\sigma^2$, i.e., $|y(n)|^2\sim \mathcal{E}(1/\sigma^2)$. Using the property that $\mathcal{E}(1/2)$ is equal to $\mathcal{X}_2^2$, we have
\begin{equation}
    \frac{2}{\sigma^2}|y(n)|^2\sim\mathcal{X}^2_2.
\end{equation}
Then, we can obtain the distribution of $\|\ybf\|_2^2$ as follows:
\begin{equation}
    \frac{2}{\sigma^2}\|\ybf\|_2^2 = \sum_{n=0}^{N-1}\frac{2}{\sigma^2}|y(n)|^2\sim\mathcal{X}^2_{2N}.
\end{equation}
Because the already-estimated $\Abf{\tilde\alphabf}$ cancels $2M$ degrees of freedom, approximately, we have
\begin{equation}
\frac{2}{\sigma^2}\|\ybf - \Abf{\tilde\alphabf}\|_2^2\sim\mathcal{X}^2_{2(N-M)}.
\end{equation}
Next, we derive the distribution of $|\abf^H_i\ybf|^2$. Because $\mathcal{CN}(0,\sigma^2)$ is rotationally invariant, $|\abf^H_i\ybf|^2$ has the same distribution as $|\abf^H_i|^2|y(0)|^2=N|y(0)|^2\sim\mathcal{E}(1/N\sigma^2)$. Thus,
\begin{equation}
    \frac{2}{N\sigma^2}|\abf^H_i\ybf|^2\sim\mathcal{X}_2^2.
\end{equation}
Then, (\ref{equ:prune_stat}) can be viewed as the quotient of two chi-square distributions:
\begin{equation}
    \frac{\frac{2}{N\sigma^2}|\abf^H_i\ybf|^2}{\frac{2}{\sigma^2}\|\ybf - \Abf{\tilde\alphabf}\|_2^2}=\frac{1}{N}\frac{|\abf^H_i\ybf|^2}{\|\ybf - \Abf{\tilde\alphabf}\|_2^2}\sim\frac{\mathcal{X}_2^2}{\mathcal{X}^2_{2(N-M)}}.
\end{equation}
Note that the statistic $\abf^H_i\ybf$ can be viewed as projecting the Gaussian statistic $\ybf$ onto a space spanned by $\abf_i$, i.e., ${\rm span}(\abf_i)$ and apparently, ${\rm span}(\abf_i)\in{\rm span}(\Abf)$. Also note that $\ybf - \Abf{\tilde\alphabf}$ can be viewed as projecting $\ybf$ onto the complement space of ${\rm span}(\Abf)$. Thus, $|\abf^H_i\ybf|^2$ and $\|\ybf - \Abf{\tilde\alphabf}\|_2^2$ are statistically independent.

Using the property that when $\mathcal{X}_2^2/2$ and $\mathcal{X}^2_{2(N-M)}/2(N-M)$ are independent,
\begin{equation}
    \frac{\mathcal{X}_2^2/2}{\mathcal{X}^2_{2(N-M)}/2(N-M)} = F_{2, 2(N-M)},
\end{equation}
we can obtain the distribution of a scaled version of (\ref{equ:prune_stat}) and it has the following form:
\begin{equation}\label{equ.xi_dist}
    \frac{N-M}{N}\frac{|\abf^H_i\ybf|^2}{\|\ybf - \Abf{\tilde\alphabf}\|_2^2}\sim F_{2, 2(N-M)}.
\end{equation}
Given the distribution of (\ref{equ:prune_stat}), the threshold can be easily obtained as
\begin{equation}
    \Xi=\frac{N}{N-M}F^{-1}_{2,2(N-M)}(1-\epsilon_a),
\end{equation}
where $F^{-1}_{2,2(N-M)}(\cdot)$ is the inverse function of the CDF of $F_{2, 2(N-M)}$. If the statistic (\ref{equ:prune_stat}) of the estimated frequency $\tilde{\omega}_i$ is smaller than $\Xi$, we prune the corresponding network node because mostly probably there is only noise at frequency $\tilde{\omega}_i$.

\begin{figure}[htbp]
  \centering
  \includegraphics[width=0.45\textwidth]{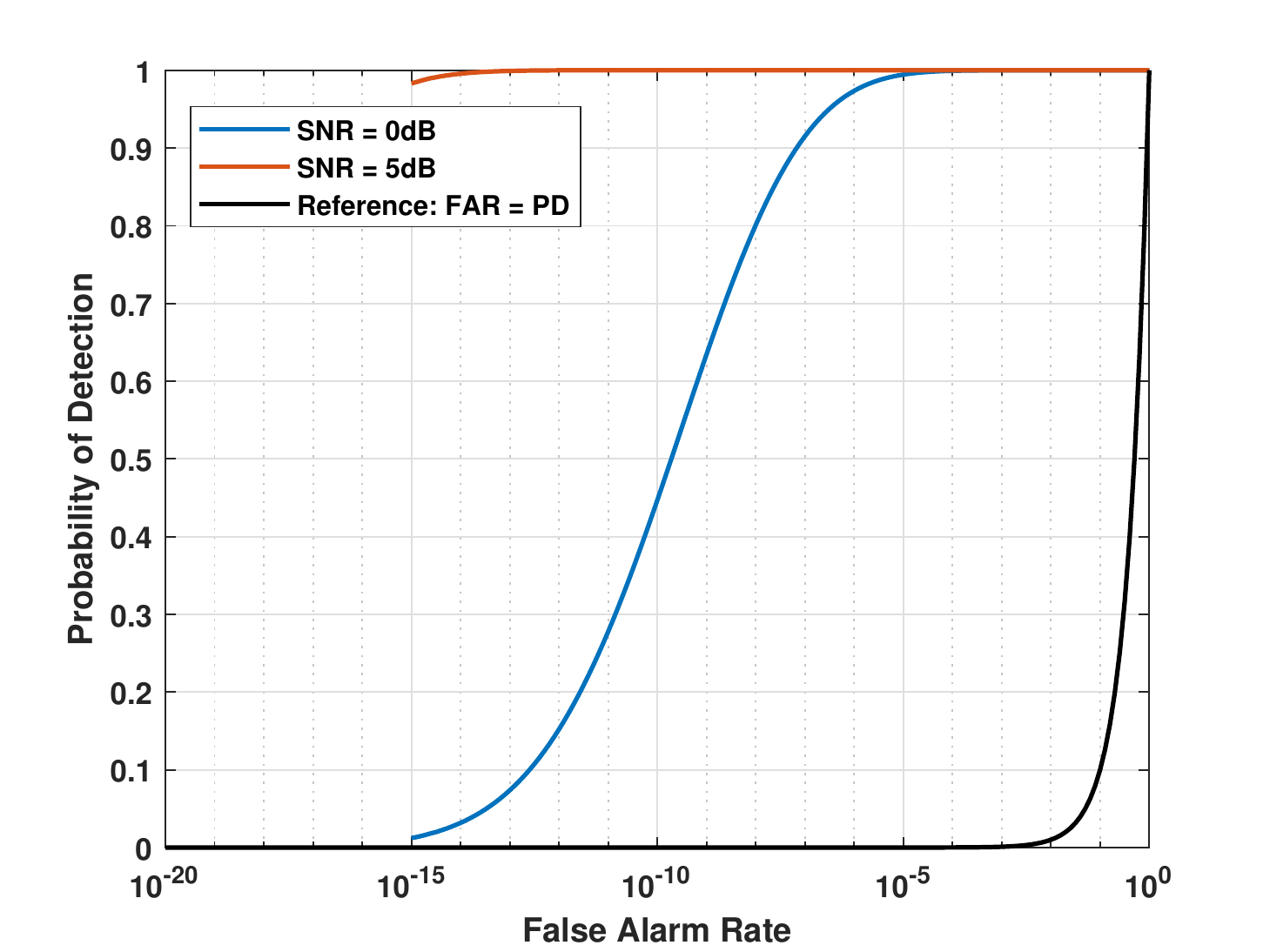}
  \caption{Theoretical ROC of the pruning method under different SNR cases}
  \label{fig.pruneROC_theory}
\end{figure}

Note that the performance of the pruning criterion can be theoretically analyzed by plotting the corresponding receiver operating characteristic (ROC) curves, for which we need to compute the false alarm rate (FAR) against the probability of detection (PD). Considering the MNN with one hidden-layer node ($M=1$), we define the PD as the probability of keeping the node when there exists a corresponding sinusoid, and define the FAR as the probability of keeping the node when the signal does not contain any sinusoid. Next, we derive PD and FAR separately. To derive FAR, we consider a signal only containing the white Gaussian noise. It is easy to know that when $\ybf = \ebf$,
\begin{equation}
    {\rm FAR} = {\rm Pr}\left(\frac{|\abf^H_i\ybf|^2}{\|\ybf - \Abf{\tilde\alphabf}\|_2^2}\geq\Xi\right) = \epsilon_a.
\end{equation}
Then we derive the PD and consider the signal containing one sinusoid with frequency $\omega_i$ and amplitude $\alpha_i$, i.e., $\ybf = \alpha_i\abf_i+\ebf$. We again derive the distribution of the numerator and denominator of $\xi$ separately first. Because the sinusoid in $\ybf$ can be subtracted by the unbiased estimate $\tilde{\alpha}_1,\tilde{\omega}_1$, we still have
\begin{equation}
    \frac{2}{\sigma^2}\|\ybf -\Abf{\tilde\alphabf}\|_2^2 =\frac{2}{\sigma^2}\|\ybf -\tilde{\alpha}_1\tilde{\abf}_i\|_2^2 \sim\mathcal{X}^2_{2(N-M)}.
\end{equation}
Next, because we have $\abf^H_i\ybf =\abf^H_i(\xbf + \ebf) =N\alpha_1 + \abf^H_i\ebf\sim\mathcal{CN}(N\alpha_1, N\sigma^2)$, $\frac{2}{N\sigma^2}|\abf^H_i\ybf|^2$ obeys the non-central chi-square distribution whose degree of freedom is $2$, and the non-centrality parameter is $2N|\alpha_1|^2/\sigma^2$. Similar to (\ref{equ.xi_dist}), the scaled version of the statistic, i.e.,
\begin{equation}
      \frac{N-M}{N}\xi \triangleq \frac{N-M}{N}\frac{|\abf^H_i\ybf|^2}{\|\ybf - \Abf{\tilde\alphabf}\|_2^2},
\end{equation}
obeys the non-central F-distribution with the degrees of freedom being $2$ and $2(N-M)$, and the non-centrality parameter is $2N|\alpha_1|^2/\sigma^2$. The probability of detection can be easily obtained by calculating ${\rm Pr}(\xi \geq \Xi)$. Denote the signal-to-noise ratio (SNR) as $10\log_{10}\frac{|\alpha_1|^2}{\sigma^2}$, the ROC curves under different SNR cases are show in Fig. \ref{fig.pruneROC_theory}. The large area under the ROC curves shows the good performance of our pruning method. More numerical analysis can be found in Section \ref{sec:simulation}.

After the merging and pruning, the corresponding weights of the reduced number of  nodes, i.e., the amplitudes and frequencies, will further be updated by the BP algorithm until convergence.

With FFT initialization and merging and pruning steps, the whole procedure of training the neural network is summarized in Algorithm \ref{Algo.1}.

\begin{algorithm}[ht]
\caption{Training of MNN}
\label{Algo.1}
\begin{algorithmic}[1]
\Require Received data sequence $\ybf$; learning rate $\gamma$, momentum parameter $\lambda$; \Ensure The optimized weight of the MNN $\tilde{\alphabf},\tilde{\omegabf}$;
\State Apply an $L$-point FFT to $\ybf$ to obtain $\ybf^f$, $L$ is commonly set as $4N$;
\State Locate the peaks of $|\ybf^f|$ as $\omega_1,..., \omega_P$, put these frequency points into the initial frequency set $\tilde{\omegabf}^0(0)$.
\State For $\omega_i, i=1,\dots, P$, also add its adjacent frequency point with larger amplitudes obtained by FFT into the initial frequency set $\tilde{\omegabf}^0(0)\in\mathbb{R}^M$.
\State Obtain the initial amplitude set $\tilde{\alphabf}^0(0)$ by using the least square method corresponding to the frequency points in $\tilde{\omegabf}^0(0)$ (see (\ref{equ:a0})).
\State $\tau=1$;
\Do
\State Initialize the NN with $\tilde{\alphabf}^{\tau-1}(0)$ and
$\tilde{\omegabf}^{\tau-1}(0)$ (when\Statex\quad\ $\tau>1$,  $\tilde{\alphabf}^{\tau-1}(0) = \tilde{\alphabf}^{\tau-1}, \tilde{\omegabf}^{\tau-1}(0) = \tilde{\omegabf}^{\tau-1}$);
\State $t = 0$;
\Do
\State Calculate $\zbf_i^{\tau}(t)$, $\abf_i^{\tau}(t)$ and $\xbf^{\tau}(t)$ by (\ref{equ.ziai})-(\ref{equx2});
\State Calculate $(\frac{\partial C}{\partial \tilde{\alphabf}^*})^{\tau}(t+1)$ and $(\frac{\partial C}{\partial \tilde{\omegabf}})^{\tau}(t+1)$ using \Statex \quad \quad \quad (\ref{equ.deralpha}) and (\ref{equ.deromega});
\State Update $\tilde{\alphabf}^{\tau}(t+1)$ and $\tilde{\omegabf}^{\tau}(t+1)$ using (\ref{equ.alphaOmegaUpdate}) and \Statex \quad \quad \quad (\ref{equ.momentum});
\State $t = t + 1$;
\State ${\bar C} =\frac{1}{N}||\ybf-\xbf^{\tau}(t)||_2^{2}$;
\doWhile{the change in ${\bar C}$ from the previous iteration is
\Statex \quad \ \ less than a pre-set tolerance $\epsilon$ (e.g., $\epsilon = 10^{-5}$)}.% <--- use \doWhile for the "while" at the end
\State {\it Merging}: If $\tilde{\omega}_j^{\tau}-\tilde{\omega}_i^{\tau}<\Delta\omega_{\min}-\sqrt{{\rm CRB}^{ij}_{\Delta}}\mathcal{N}^{-1}(\epsilon_f)$,
\Statex \quad  \ \ merge them into one, average their frequencies to be
\Statex \quad  \ \ $(\tilde{\omega}_i^{\tau}+\tilde{\omega}_j^{\tau})/2$, and combine the amplitudes into $\tilde{\alpha}_i^{\tau}+\tilde{\alpha}_j^{\tau}$.
\State {\it Pruning}: If $\xi$ of $\tilde{\omega}_i$ is smaller than $\Xi$, prune the
\Statex \quad  \ \  corresponding node.
\State Taking modulo $\tilde{\omega}_i^{\tau}\!\leftarrow\!\text{mod}(\tilde{\omega}_i^{\tau}, 2\pi)$.
\State Result: $\tilde{\bm\omega}^{\tau}, \tilde{\bm\alpha}^{\tau}$ after merging and pruning.
\State $\tau=\tau+1$.
\doWhile{no sinusoidal components are merged or pruned.}
\end{algorithmic}
\end{algorithm}
%{\em Step 1}, restrict $\tilde{\omega}_i$ to $\left[0, 2\pi\right]$. Utilizing the fact that $\hat{y}$ is a function of $\tilde{\omega}_i$ with period $2\pi$, we have
%\ben
%\tilde{\omega}_i \longleftarrow \text{mod}(\tilde{\omega}_i, 2\pi), i = 1,2,\dots, M
%\een
%
%{\em Step 2}, change the interval $\tilde{\omega}_i\in\left[0, 2\pi\right]$ into $\tilde{\omega}_i\in\left[-\pi, \pi\right]$,i.e.,
%\ben
%\begin{split}
%\tilde{\omega}_i = \left\{\ba{ll}
%\tilde{\omega}_i - 2\pi,  & \tilde{\omega}_i \in \left[\pi, 2\pi\right], \\
%\tilde{\omega}_i,  & \tilde{\omega}_i \in \left[0,\pi\right].
%\ea
%\right. i = 1,2,\dots, M.
%\end{split}
%\een

%After training the NN and removing $\tilde{\omega}_i$'s ambiguity, the $M$ pairs of $\{\tilde{\omega}_i,\tilde{\alpha}_i\}$ contain complete spectrum information of $\ybf$.

\subsection{Complexity Analysis}
From Line 10 - 12 in Algorithm \ref{Algo.1}, (\ref{equ.ziai}) needs $MN$ multiplications. The complexity of (\ref{equx2}) and (\ref{equ.deralpha}) are both ${\cal O}(MN)$. In (\ref{equ.deromega}), $\nbf\odot(\ybf- \xbf)^*$ needs $N$ multiplications, the complexity of multiplying $\Abf^T$ and $[\nbf\odot(\ybf- \xbf)^*]$ is ${\cal O}(MN)$, and the complexity of the element-wise multiplication between $2\tilde{\alphabf}$ and  $\left[\Abf^T[\nbf\odot(\ybf- \xbf)^*]\right]$ is ${\cal O}(2M)$. Thus, the total computational complexity of (\ref{equ.deromega}) is ${\cal O}(2M+N+MN)$. Finally, (\ref{equ.alphaOmegaUpdate}) and (\ref{equ.momentum}) need $2M$ and $4M$ multiplications, respectively. Thus, the whole process takes ${\cal O}(I(4MN+8M+N))$ with $I$ being the total number of iterations. Moreover, it can be envisioned that the neural network-like structure of the MNN allows for ultra-efficient implementation of parallel computation conducted on a GPU, which is out of the scope of this paper and is left to future investigation.

\section{Numerical Simulation}\label{sec:simulation}
In this section, we provide several simulation examples to verify the effectiveness of the line spectral estimation using the MNN. For all the cases, the SNR is defined as follows:
\begin{equation}
    {\rm SNR} = 10\log_{10}\frac{\|\xbf\|_2^2}{\|\ebf\|_2^2}{\rm (dB)}.
\end{equation}

\subsection{Comparison of Estimation Precision}
\begin{figure}[htb]
\centering
\includegraphics[width=3.2in]{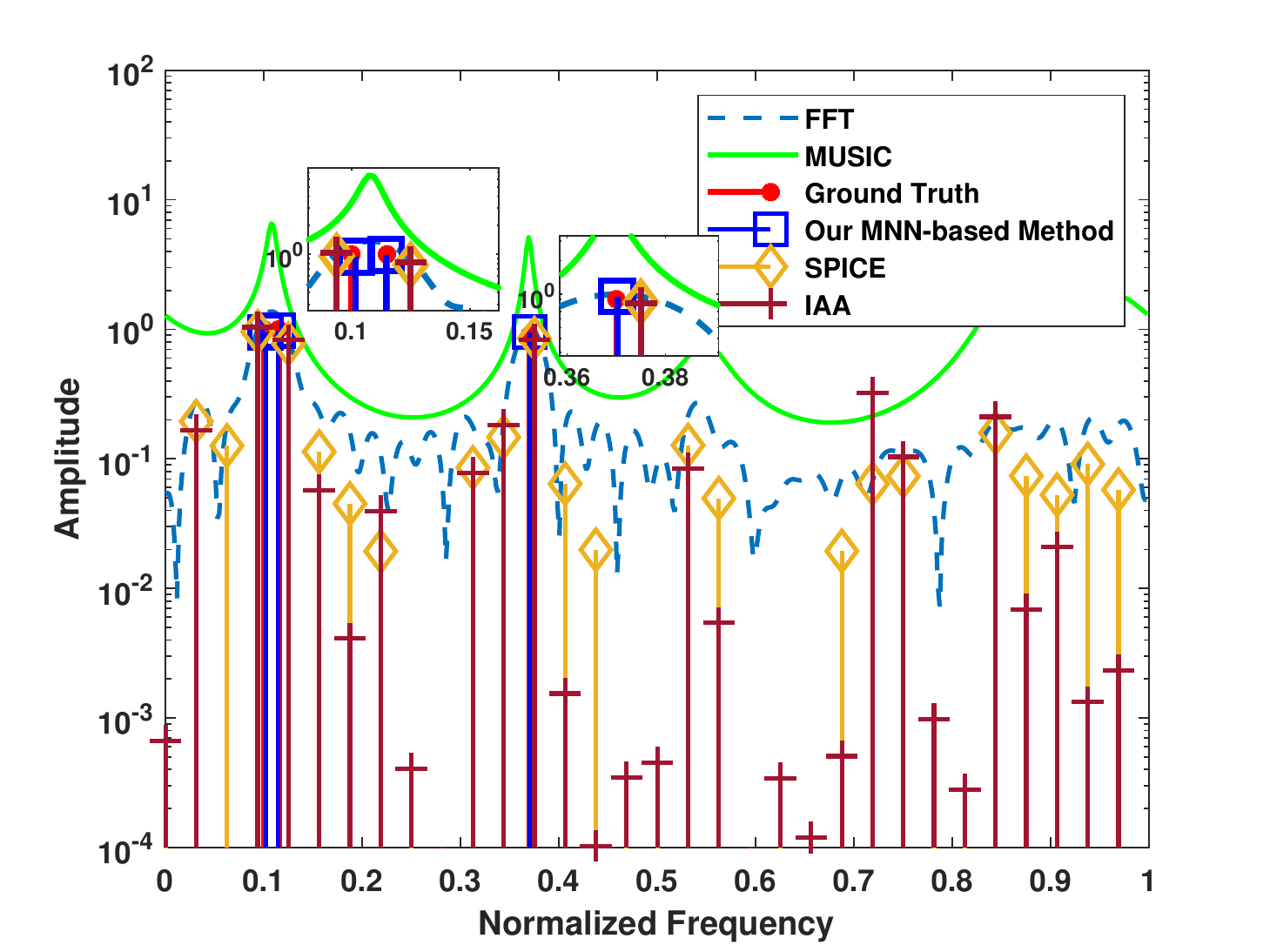}
\caption{Spectrum estimation results of different algorithms when SNR is 10 dB with $M=6$.}
\label{fig:spectral30db}
\end{figure}

We first compare the performance of different spectral estimation methods including our MNN and several other widely-used methods. We simulate a $N=32$ point signal which contains $K=3$ complex sinusoidal components with normalized digital frequency $0.1, 0.115$ and $0.37$. The signal is contaminated by zero-mean white Gaussian noise, and the signal-to-noise ratio (SNR) is 10dB. The amplitudes of the signal are marked by red dots in Fig. \ref{fig:spectral30db}. For comparison, we also provide the results of other spectral estimation algorithms, including the FFT, the MUSIC and two grid-based methods, i.e., SPICE and IAA \cite{SZL14, SPP20}. The number of the grid points in the frequency domain used by the grid-based methods is equal to the signal length, i.e., $32$. Fig. \ref{fig:spectral30db} shows the spectral estimation results obtained by different algorithms with $M=6$. It is clear that FFT and MUSIC cannot distinguish the two complex sinusoids with frequencies $0.1$ and $0.115$. Although the grid-based methods can distinguish these two signals, their performance are limited by the granularity of the grids. Our MNN-based method has higher estimation accuracy of the frequencies and amplitudes of all three cosine waves. Moreover, only our MNN-based method does not need to know the number of sinusoidal signals before-hand. When we assume that $M=6$ instead of $3$, the extra components can be merged and pruned by our MNN-based method and the number of sinusoidal components is correctly estimated, which means lower false-alarm probability and no need to know exact $K$ in advance.

%\begin{figure}[htbp]
%  \centering
%  \subfloat[]{\includegraphics[width=0.45\textwidth]{spectral30db_3.eps}}\\
%  \subfloat[]{\includegraphics[width=0.45\textwidth]{spectral30db_5.eps}}\\
%  \caption{Spectrum estimation results of different algorithms when SNR is 30 dB with a) $M=3$, b) $M=5$.}\label{fig:spectral30db}
%\end{figure}

\begin{figure}[htbp]
  \centering
  \subfloat[]{\includegraphics[width=0.45\textwidth]{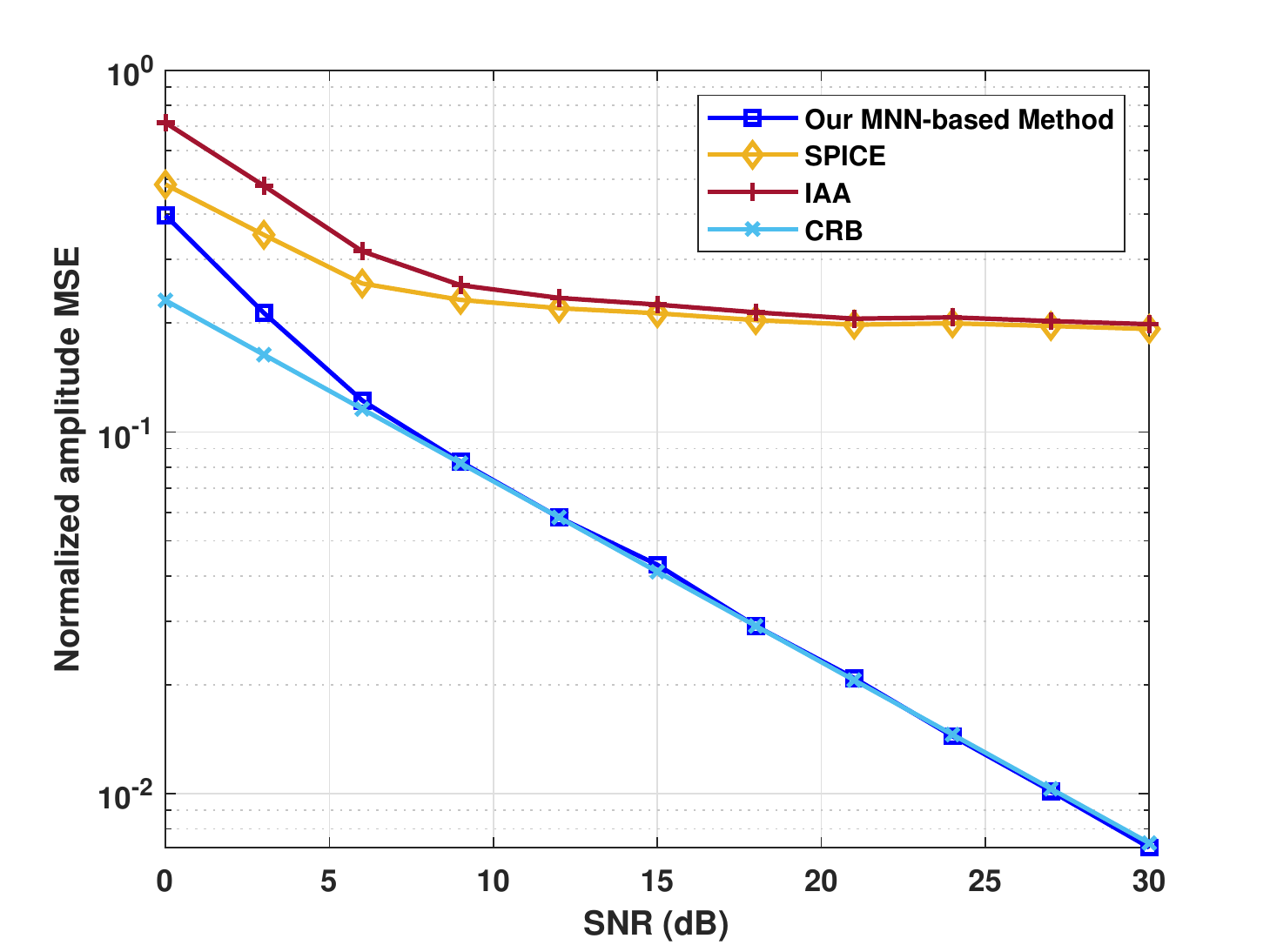}}\\
  \subfloat[]{\includegraphics[width=0.45\textwidth]{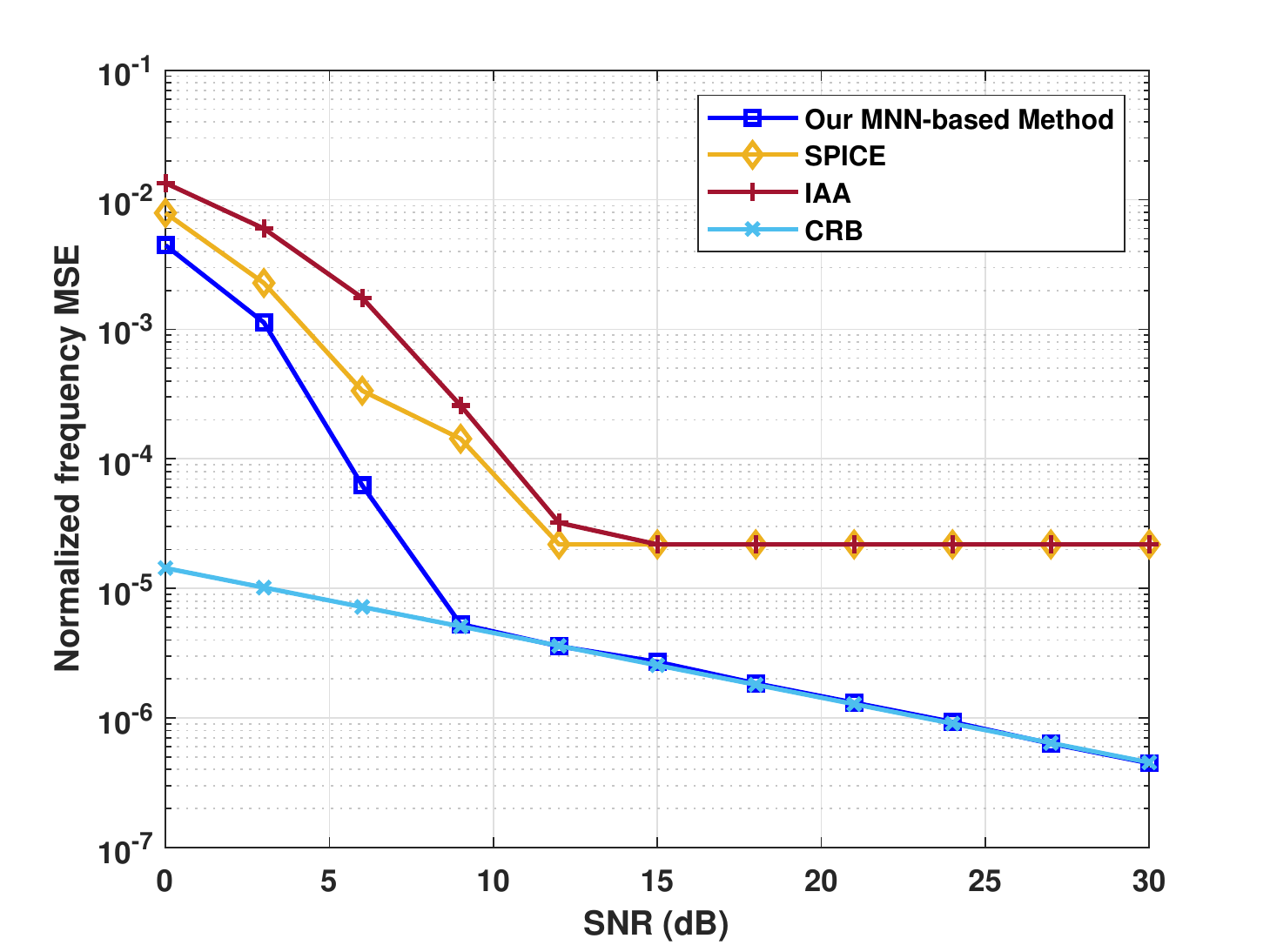}}\\
  \caption{Normalized a) amplitude and b) frequency MSE versus SNR for the parameter estimation problem.}\label{fig:MSE}
\end{figure}

We then compare the Cram\'er–Rao bound (CRB) and the normalized mean square error (MSE) of amplitudes and frequencies obtained by different algorithms under different SNR cases. Under each SNR case, we adopt a 1000 times Monte Carlo simulation. We set a 32-point time series which contains three complex sinusoidal components with the digital angular frequencies at $2\pi\times[0.1, 0.22, 0.37]$ and the amplitudes are randomly generated in each Monte Carlo run. Fig. \ref{fig:MSE} shows that when SNR increases, the performance of SPICE and IAA\cite{SZL14, SPP20} will not improve. It is because the performance of the grid-based methods are greatly limited by the not-good-enough grid, while our MNN-based method eliminates this problem and outperforms its four counterparts. Additionally, when SNR increases, only the MSE of our MNN-based method come close to the CRB.
%The other two blue circles with small amplitude (at frequencies of 0.27 and 0.48) are due to the fact that we do not know the number of signals from the beginning, so we set up five hidden layer nodes, but because they are more than the actual. Finally, the signal amplitude corresponding to the remaining two nodes converges to a very small value, so it can be judged that the number of actual signals is 3.
\subsection{Validation of Merging and Pruning}
In this subsection, we validate the performance of our merging and pruning method. We first investigate the performance of the merging and pruning method separately by plotting the corresponding ROC curves. The necessary PD and FAR values are obtained by 1000 Monte Carlo trials.

\begin{figure}[htbp]
  \centering
  \includegraphics[width=0.45\textwidth]{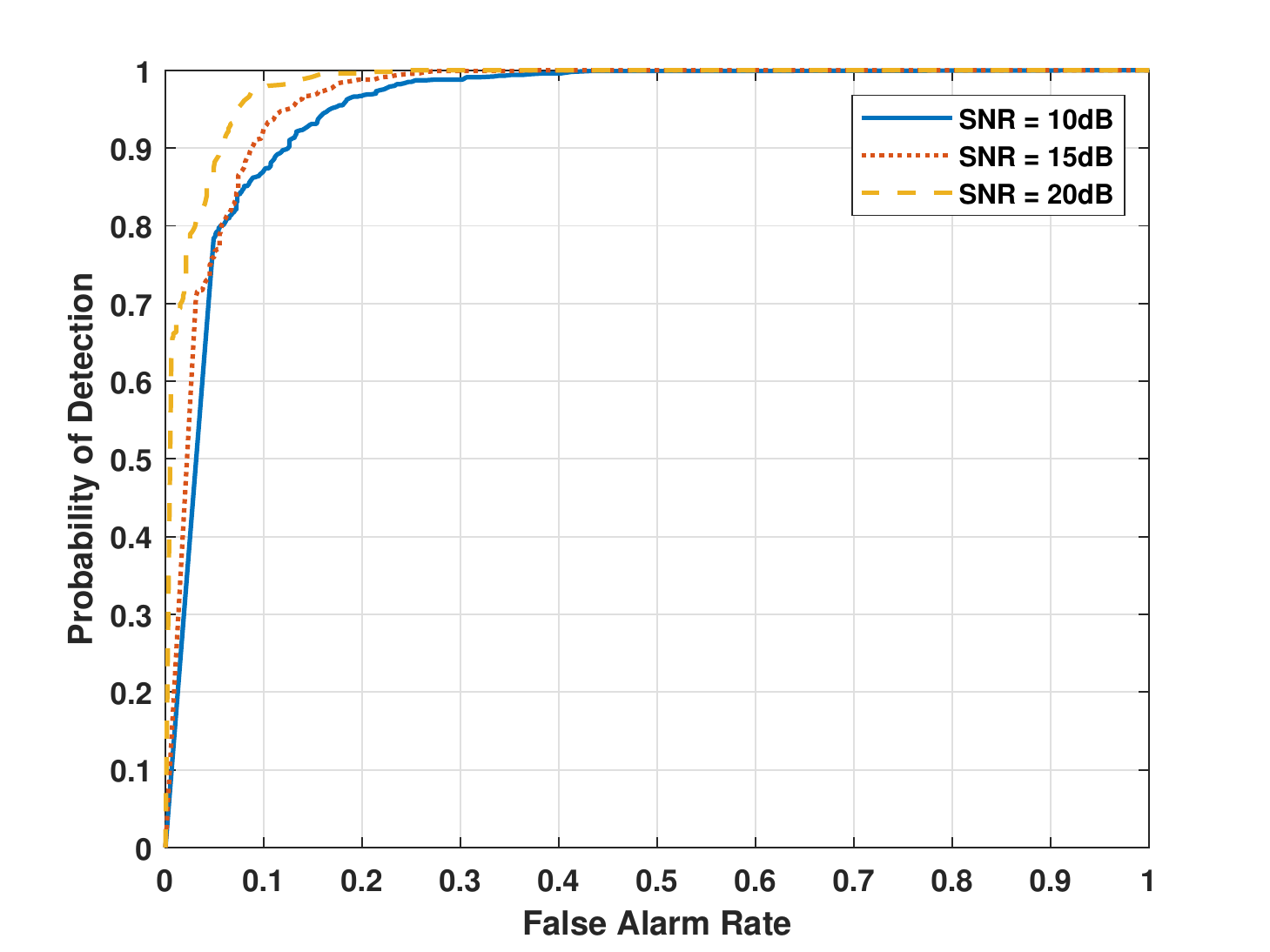}
  \caption{ROC curve of the merging criterion under different SNR cases.}\label{fig.mergeROC}
\end{figure}

To plot the ROC curve of the merging criterion, we consider a MNN with two hidden-layer nodes corresponding to two close frequencies and see if they will be merged by our method. The threshold $\Delta\omega_{\min}$ is set as zero for all the cases. The PD of the ROC curve is defined as the probability of correctly keeping the two nodes when there are two sinusoids, and the FAR is defined as the probability of erroneously keeping the two nodes when there is only one sinusoid. To numerically obtain PD, we simulate the $32$-point input signal which contains two complex sinusoidal components with the frequency $[0.5, 0.5+\frac{1}{16N}]\times2\pi$ and the same power level. Note that the two sinusoids are extremely close in the frequency domain and the frequency difference is much smaller than the resolution of FFT. We initialized the corresponding MNN with two hidden-layer nodes by the method in the Section \ref{sec.initial}. To obtain FAR, we simulate the input signal which contains only one sinusoid with the frequency $0.5\times2\pi$, and the MNN is also initialized with two hidden-layer nodes. By varying $\epsilon_f$ from $0$ to $1$, we can plot the ROC curve under different SNR cases (shown in Fig. \ref{fig.mergeROC}). It is clear that the area under the ROC curves is close to one, which shows the good performance of our merging criterion. Also, when SNR increases, our merging criterion performs better, which conforms our intuition.

\begin{figure}[htbp]
  \centering
  \includegraphics[width=0.45\textwidth]{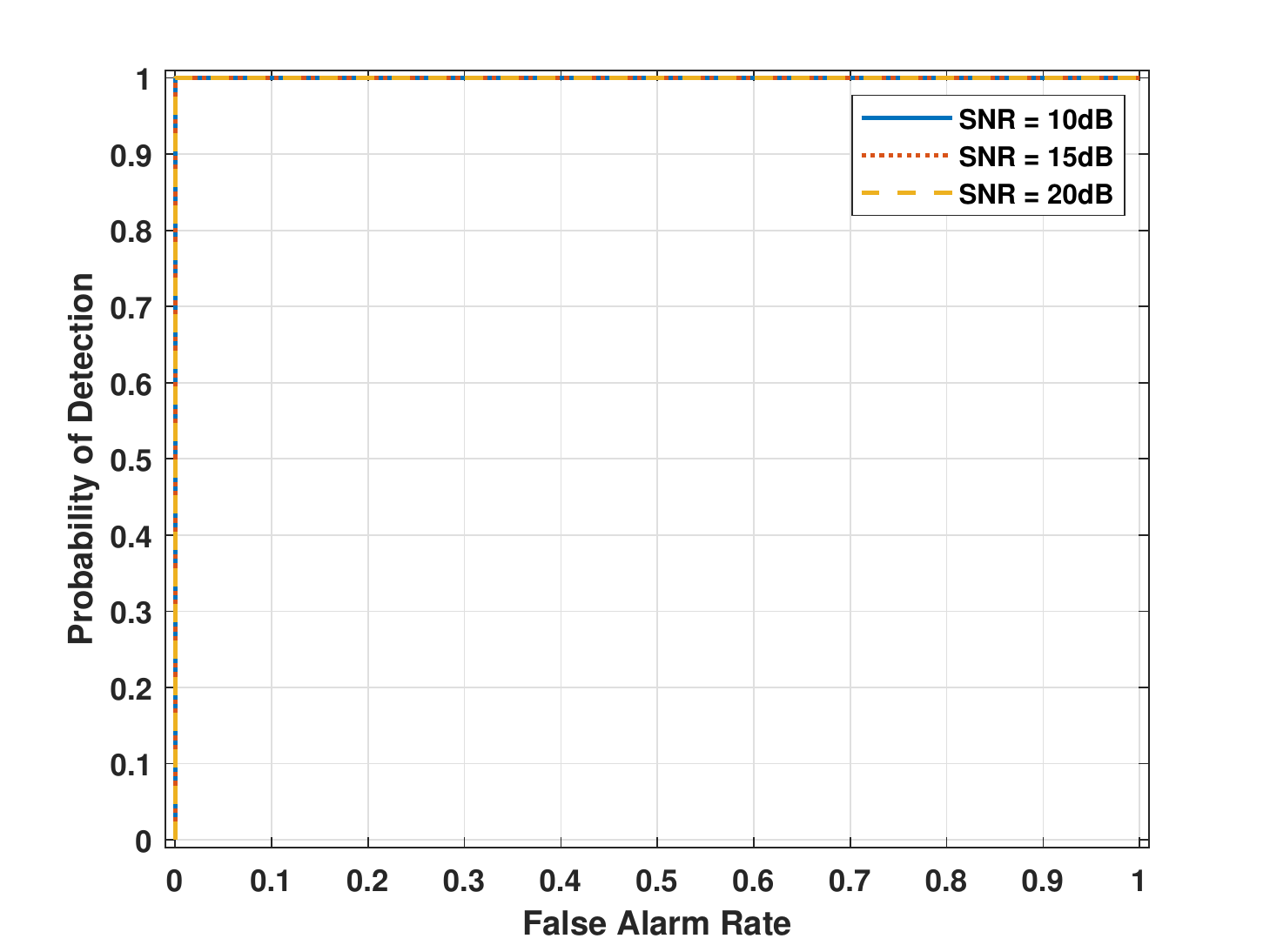}
  \caption{ROC curve of pruning the only one hidden-layer node under different SNR cases}\label{fig.pruneROC1}
\end{figure}

Next, we investigate the performance of the pruning method. We consider two different scenarios. We first consider the scenario that the MNN has only one hidden-layer nodes and see if it will be pruned by our pruning methods. In this scenario, we define the PD as the probability of correctly keeping the nodes when there is a sinusoid, and define the FAR as the probability of erroneously keeping the node when the signal only contains the white Gaussian noise. To obtain the PD, we simulate the $32$-point time series which contains a sinusoid with the normalized frequency $0.5$. To obtain FAR, we simulate the time series which is just the white Gaussian noise. The MNN is initialized with only one hidden-layer node, and PD and FAR are obtained by Monte Carlo trials. By varying $\epsilon_a$ from $0$ to $1$, we can plot the ROC curves under different SNR cases (see Fig. \ref{fig.pruneROC1}). Note that in this scenario, SNR is only defined for the signal with one sinusoid. Fig. \ref{fig.pruneROC1} shows that the area under the ROC curves is $1$, which means the extremely good performance of our pruning method when SNR $>10$dB. This conforms to our theoretically analysis in Section \ref{sec.prunecriterion}.

\begin{figure}[htbp]
  \centering
  \includegraphics[width=0.45\textwidth]{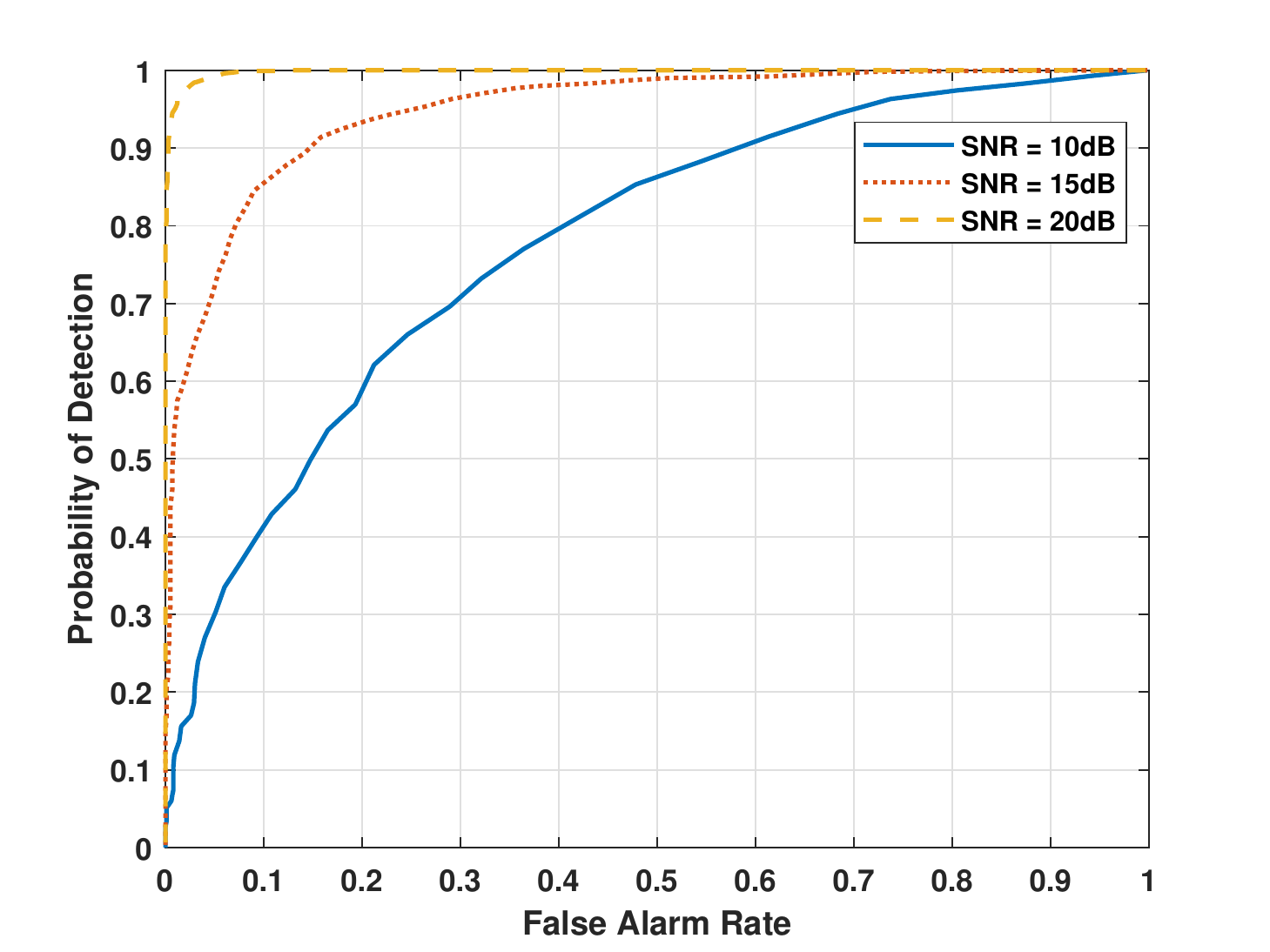}
  \caption{ROC curve of pruning the hidden-layer node corresponding to the weaker sinusoid under different SNR cases}\label{fig.pruneROC}
\end{figure}

We next consider the scenario that the MNN has two hidden-layer nodes and investigate if the node corresponding to the weaker sinusoidal component will be pruned. We define the PD as the probability of correctly keeping the two nodes when there are two sinusoids, and define the FAR as the probability of erroneously keeping the node corresponding to the weaker sinusoid when is only one sinusoid. To obtain PD, we simulate a $32$-point signal with two sinusoids. The normalized frequencies $\omega_1, \omega_2$ are $0.5, 0.8$, respectively, and the absolute amplitudes $|\alpha_1|, |\alpha_2|$ are $1, 0.1$, respectively. Note that the second sinusoid is much weaker than the first. To obtain FAR, we simulate the signal which contains one sinusoid with the frequency $\omega_1 = 0.5\times2\pi$ and the absolute amplitude $|\alpha_1|=1$. The MNN is initialized with two hidden-layer nodes, and the PD and the FAR are obtained by Monte Carlo trials. By varying $\epsilon_a$, we can plot the ROC curves under different SNR cases (shown in Fig. \ref{fig.pruneROC}). It shows than even when $|\alpha_2|\ll|\alpha_1|$, our pruning method can have good performance when SNR $>10$dB.

\begin{figure}[htbp]
  \centering
  \includegraphics[width=0.45\textwidth]{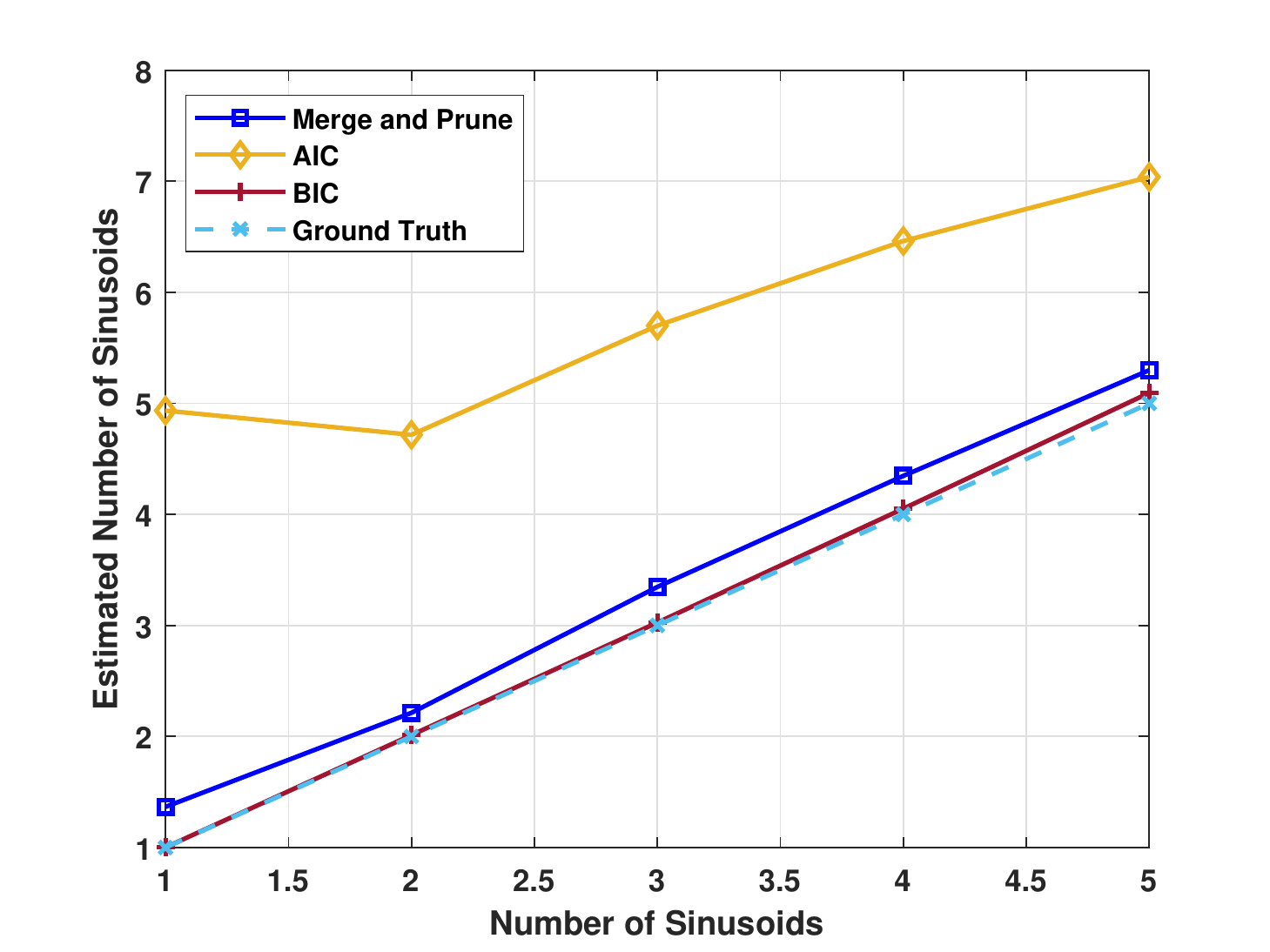}
  \caption{Results of different methods of estimating the number of sinusoids.}\label{fig:model_order}
\end{figure}

To investigate the performance of our merging and pruning method when they are used together, we consider different cases with the number of sinusoids varying from $1$ to $5$, and under each case, we adopt a $1000$ times Monte Carlo run. We keep $N=32, \epsilon_f =1\times10^{-6}, \epsilon_a=1\times10^{-6}$ and SNR $=10$dB in each Monte Carlo run. We use two widely-used model-order selection methods, i.e., AIC and BIC \cite{SS2004}, as benchmarks. Specifically, we assume different model-orders and design the MNN with different numbers of hidden-layered nodes. We train these MNNs and substitute the results into the corresponding AIC and BIC metrics. The estimated model-order is the one that minimizes the AIC or BIC metric. The results are shown in Fig. \ref{fig:model_order}. It is clear that the performance of AIC is the inferior one among all the methods and BIC slightly outperforms our merging and pruning method. But our method is computationally much more efficient that the BIC, which needs to try different number of sinusoids before finding the one minimizing the BIC metric.

\subsection{Convergence Performance}
\begin{figure}[htbp]
  \centering
  \includegraphics[width=0.45\textwidth]{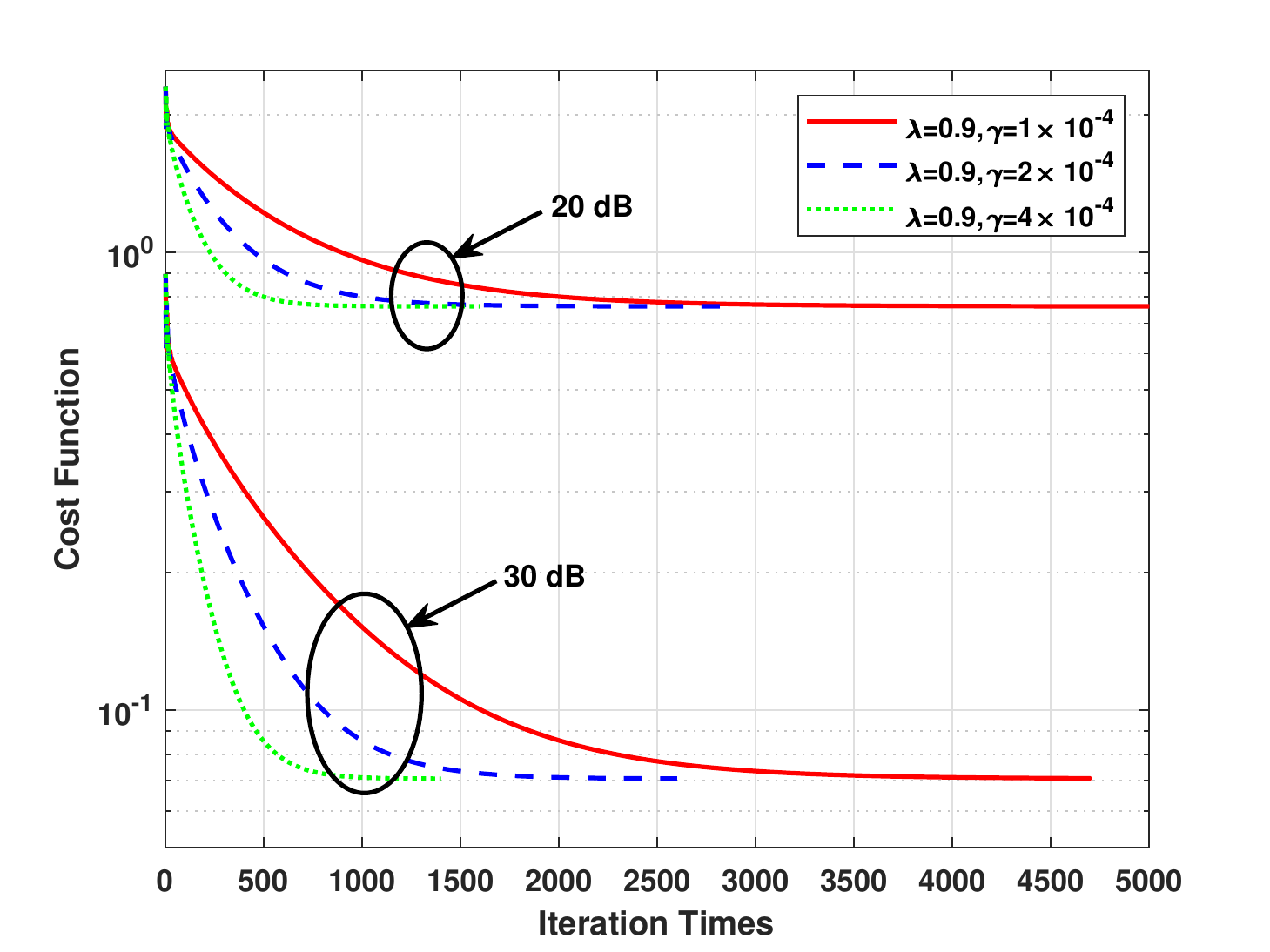}
  \caption{The variation of the cost function value during the iterations with different learning rates and fixed momentum parameter under different SNR cases.}\label{fig:convergence1}
\end{figure}

To investigate the convergence property of our MNN-based method for the non-convex spectral estimation problem, we also show the variation of the cost function value during the iterations with different learning rates $\gamma$ (see Fig. \ref{fig:convergence1}) under different SNR cases. It is clear that our MNN-based method can always achieve the convergence with different learning rates. When fixing the momentum parameter, larger learning rate provides faster convergence, which conforms our intuition.

%\begin{figure}[htbp]
%  \centering
%  \includegraphics[width=0.45\textwidth]{mnn_converge.eps}
%  \subfloat[]{\includegraphics[width=0.4\textwidth]{learning_rate.eps}}\\
%  \subfloat[]{\includegraphics[width=0.4\textwidth]{momentum_rate.eps}}
%  \caption{The variation of the cost function value during the iterations with different learning rate and fixed momentum parameter under different SNR cases.}\label{fig:convergence}
%\end{figure}

\begin{figure}[htbp]
  \centering
  \includegraphics[width=0.45\textwidth]{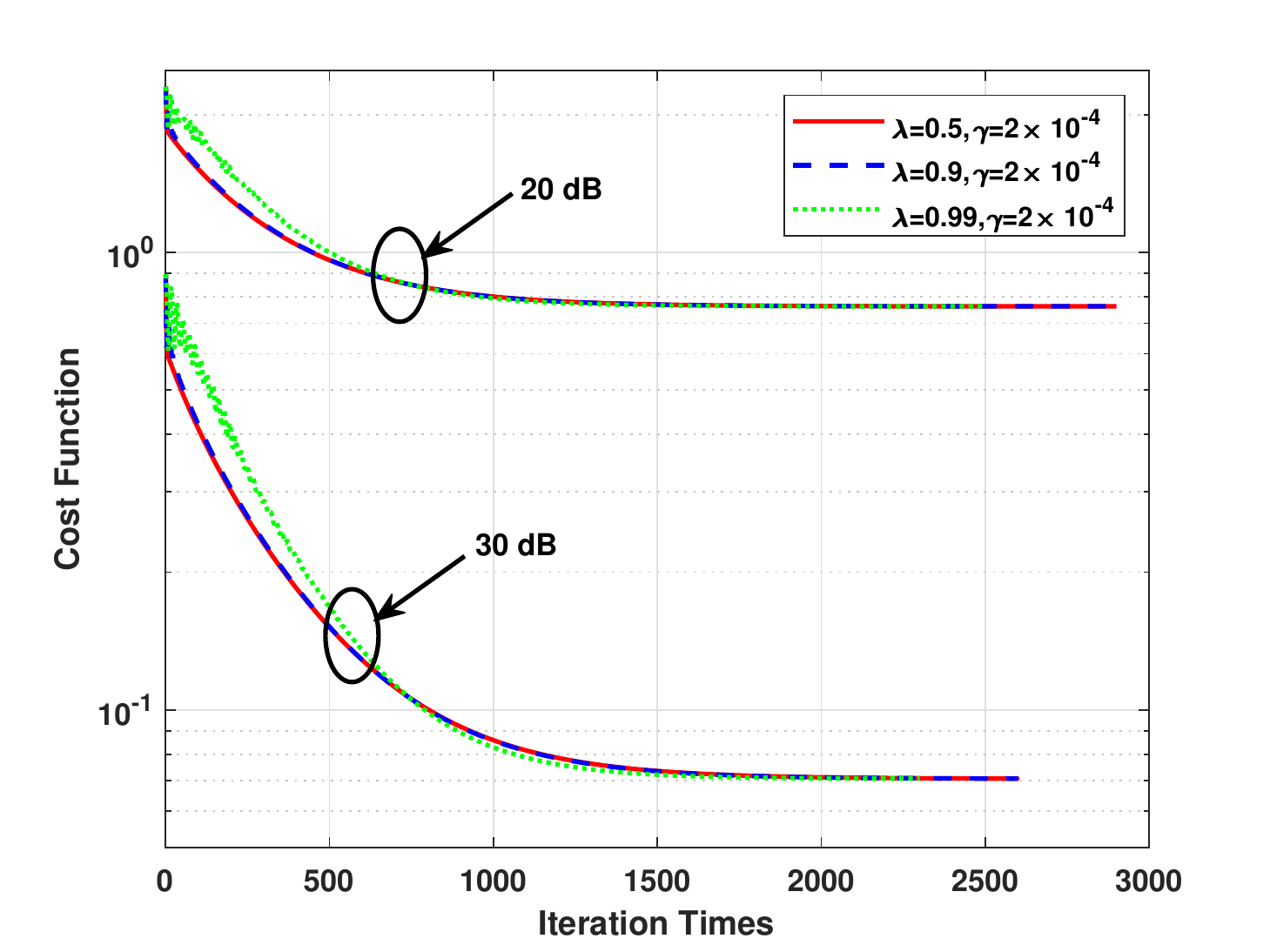}
  \caption{The variation of the cost function value during the iterations with fixed learning rate and varying momentum parameters under different SNR cases.}\label{fig:convergence2}
\end{figure}

We also show the variation of the cost function value with with  fixed  learning rate  and varying momentum parameters (see Fig. \ref{fig:convergence2}). When using different momentum parameters, our method can also achieve the convergence. We can also find that the the cost function will descend more smoothly with small momentum parameters, while large  momentum parameters may sometimes lead to the increase of the cost function during the iterations, and that is why the momentum method can help to jump out of the local minima.

\subsection{A Challenging Case}
\begin{figure}[htb]
\centering
\includegraphics[width=3.2in]{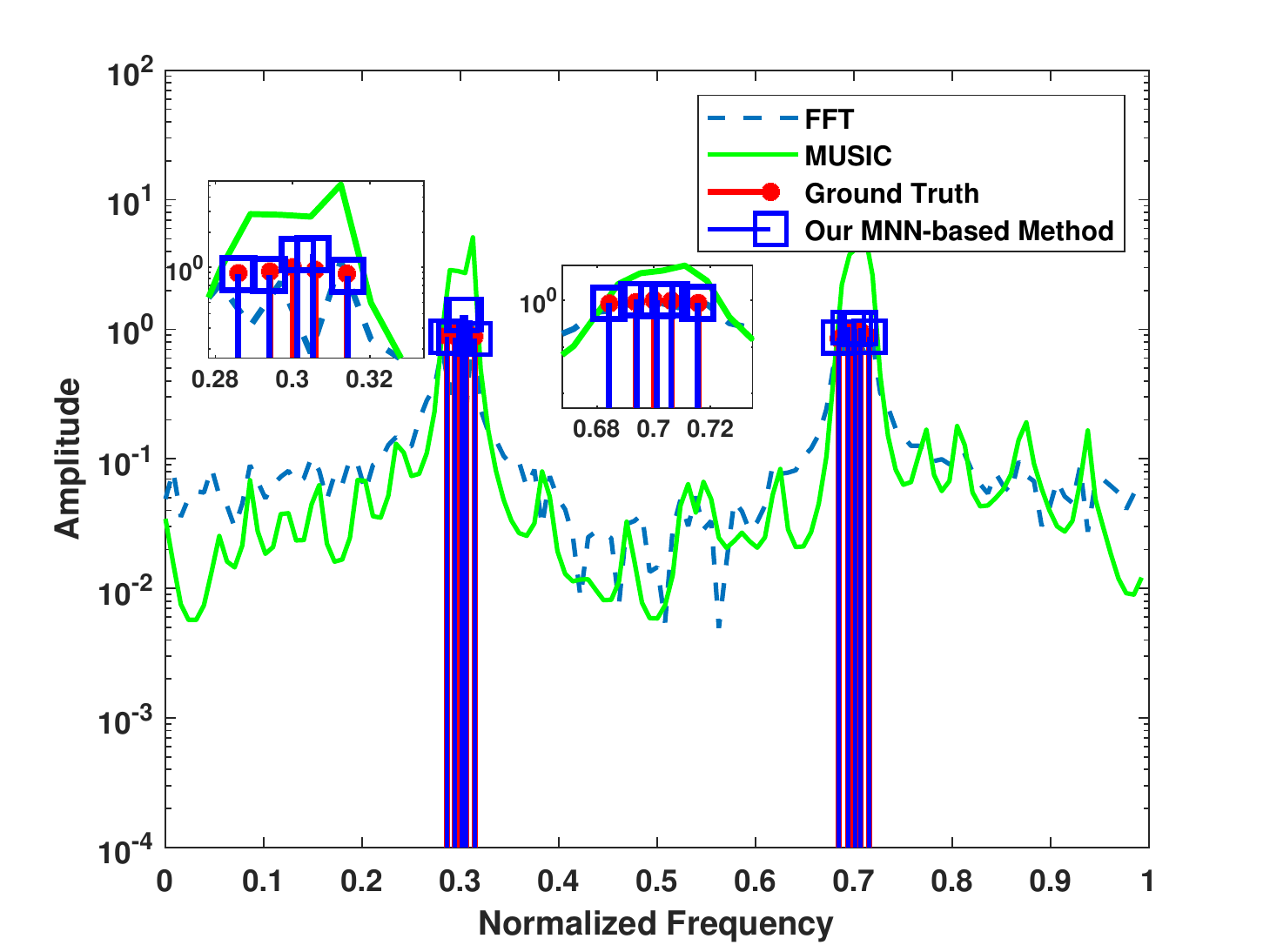}
\caption{Spectrum estimation results of different algorithms of the signal with two clusters when SNR $=20$ dB. }
\label{fig:clusterEst}
\end{figure}

In the last example we simulate the more challenging scenario where there are clusters of closely-spaced sinousoids. % are cluster complicated application of our MNN-based method, i.e., the cluster estimation problem. Cluster estimation is a challenging problem in many signal processing fields like radar target detection and communication channel estimation due to the multitude of the model order and the very close frequencies of different sinusoids.
We consider a $N=128$ time series with two clusters. Each cluster contains five sinusoids with the digital angular frequencies at $2\pi\times[0.3, 0.3+0.75/N, 0.3-0.75/N, 0.3+1.8/N, 0.3-1.8/N]$ and $2\pi\times [0.7, 0.7+0.8/N, 0.7-0.8/N, 0.7+2/N, 0.7-2/N]$. The amplitudes of the sinusoids are shown in Fig. \ref{fig:clusterEst}, and the SNR of the input signal is $20$ dB. Note that to estimate the close sinusoids in frequency domain of each cluster, we initialize the MNN with the frequency $\tilde{\omega}_i, \tilde{\omega}_i+\frac{2\pi}{L}, \tilde{\omega}_i-\frac{2\pi}{L}$, where $\tilde{\omega}_i, i=1,\dots,4$ denote the FFT spectrum peaks as mentioned in Section \ref{sec.initial}. Thus, the initialized number of the hidden-layered nodes is $12$ instead of the true value $10$. Fig. \ref{fig:clusterEst} shows that our MNN-based method can correctly estimate all the sinusoids in the two clusters with relatively low estimation error. Also, by using the merging and pruning method, the model order is correctly estimated. The performance of MNN on this complicated scenario shows its validity and wide applications in the future.

%\begin{figure}[ht]
%\centering
%\begin{tabular}{c}
%{\psfig{figure=spectral_analysis.eps ,width=3.4in}}
%\end{tabular}
%\caption{Output SNR of the relay network achieved by the NAF-BP as the number of iterations  with respect to different number of the relay nodes and the relay antennas.} \label{varyMrN}
%\end{figure}

%\begin{figure}[ht]
%\centering
%\begin{tabular}{c}
%{\psfig{figure=spectral_analysi_converg.eps ,width=3.4in}}
%\end{tabular}
%\caption{Output SNR of the relay network achieved by the NAF-BP as the number of iterations  with respect to different number of the relay nodes and the relay antennas.} \label{varyMrN}
%\end{figure}

\section{Conclusions} \label{sec:con}
In this paper, we present a novel signal modeling tool named model-based neural network (MNN) and solve the classic line spectral estimation problem as a showcase of MNN. By choosing the complex exponential function as the activation function and reviewing the complex amplitude and digital angular frequency as the network weights, we model the signal by a three-layered neural network and use the back-propagation (BP)  algorithm to train this network. To overcome the non-convexity of the line spectral estimation problem, we use the momentum method to jump out of the local minima and use FFT to obtain a good initialization. To determine the number of sinusoids in the signal, we also artfully design the rules of merging and pruning the hidden-layer nodes of MNN. %This novel model-order selection method avoids the exhaustive search of many existing selection methods and has much lower computational complexity.
The simulations show that our proposed method is performance-wise optimal and computation-wise simple compared to many existing and widely-used spectral estimation methods.

% Owing to the layered structure, the MNN can also be optimized using the back-propagation (BP) algorithm based on the chain rule of derivative; thus, the MNN allows for parallel computation conducted on a graphic processing unit (GPU) ultra-efficiently, which makes it suitable for solving large-scale problems.

% We believe that the model-based neural network is a powerful modeling tool that can find other applications in the future.

% The MNN was actually used for distributed relay beamforming in our previous work \cite{WangJiang2020} (under the name ``quasi-neural network''), which drew analogies between a relay network and the ANN. We believe that the concept of MNN will inspire future researches beyond the spectral estimation and relay communications.

%By choosing the complex exponential function as the activation function and reviewing the complex amplitude and digital angular frequency as the network weights, we transform the traditional neural network into a modeling tool. BP algorithm is used to train this NN and FFT is used to provide a good initial for this non-convex problem. Numerical examples show that our proposed method outperforms several existing spectral methods because it has high frequency resolution and its results are not confined to the grid points.

% \appendix
\section*{Appendix: The derivation of (\ref{equ.CRBomega})}
According to \cite{stoica2005spectral}, the Fisher Information Matrix (FIM) of (\ref{equ.2sin}) is
\begin{equation}
    {\rm FIM}^{ij} = \frac{2}{\sigma^2}{\rm Re}\begin{bmatrix}
    \Pi_{ii} & \Pi_{ij} \\
    \Pi_{ji} & \Pi_{jj}
    \end{bmatrix},
\end{equation}
where
\begin{equation}
    \begin{split}
        &\Pi_{ii} = \left(\frac{\partial\ybf}{\partial\omega_i}\right)^H\left(\frac{\partial\ybf}{\partial\omega_i}\right) = |\alpha_i|^2\sum_{n=0}^{N-1}n^2,\\
        &\Pi_{jj} = \left(\frac{\partial\ybf}{\partial\omega_j}\right)^H\left(\frac{\partial\ybf}{\partial\omega_j}\right) = |\alpha_j|^2\sum_{n=0}^{N-1}n^2,\\
        &\Pi_{ij} = \left(\frac{\partial\ybf}{\partial\omega_i}\right)^H\left(\frac{\partial\ybf}{\partial\omega_j}\right) = \alpha_i^*\alpha_j\sum_{n=0}^{N-1}n^2e^{j(\omega_j - \omega_i)n},\\
        &\Pi_{ji} = \left(\frac{\partial\ybf}{\partial\omega_j}\right)^H\left(\frac{\partial\ybf}{\partial\omega_i}\right) = \alpha_i\alpha_j^*\sum_{n=0}^{N-1}n^2e^{j(\omega_i - \omega_j)n}.
    \end{split}
\end{equation}
Then
\begin{equation}
    \begin{split}
        &{\rm Re}[\Pi_{ii}] = |\alpha_i|^2\rho_1,\\
        &{\rm Re}[\Pi_{jj}] = |\alpha_j|^2\rho_1,\\
        &{\rm Re}[\Pi_{ij}] = {\rm Re}[C_{ji}] = {\rm Re}\left[\alpha_i^*\alpha_j\rho_2\right].
    \end{split}
\end{equation}
with $\rho_1 = \sum_{n=0}^{N-1} n^2$ and $\rho_2 = \sum_{n=0}^{N-1}n^2e^{j(\omega_j - \omega_i)n}$.
Thus, the CRB matrix of (\ref{equ.2sin}) is
\begin{equation}
\begin{split}
     &{\rm CRB}^{ij} = [{\rm FIM}^{ij}]^{-1} \\
     &= \frac{\sigma^2}{2}\frac{1}{{\rm Re}[\Pi_{ii}]{\rm Re}[\Pi_{jj}] - {\rm Re}[\Pi_{ij}]^2}\begin{bmatrix}
    {\rm Re}[\Pi_{jj}] & -{\rm Re}[\Pi_{ij}] \\
    -{\rm Re}[\Pi_{ji}] & {\rm Re}[\Pi_{ii}]
    \end{bmatrix}.
\end{split}
\end{equation}

\bibliographystyle{ieeetr}
\bibliography{all}

\end{document}